\documentclass[]{article}

\usepackage{cite}

\def\bbox#1{\mbox{\boldmath$#1$}}

\def\trunc#1{[\mkern - 2.5 mu  [#1] \mkern - 2.5 mu ]}
\def\corresponds{{\lower.2ex\hbox{=}}{\rm\kern-.75em^\triangle}}
\def\succsim{\succ\kern-.9em_\sim\kern.3em}
\def\precsim{\prec\kern-1em_\sim\kern.3em}
\def\slantfrac#1#2{\kern1em^{#1}\kern-.3em/\kern-.1em_{#2}}
\def\lfrac#1#2{{}^{#1\!}\kern-.0em/_{#2}}
\def\trunc#1{[\mkern - 2.5 mu  [#1] \mkern - 2.5 mu ]}

\begin{document}

\bibliographystyle{myprsty}

\vspace{2cm}

\noindent
Preprint hep-ph/0005198\\
MPIPKS-Preprint mpi-pks/0005002\\

\vspace{2cm} 

\begin{center}
{\Large \sf Asymptotic Improvement of}\\[2ex]
{\Large \sf Resummations and Perturbative Predictions}\\[2ex]
{\Large \sf in Quantum Field Theory}
\end{center}

\vspace{1cm}

\begin{center}
U.~D.~Jentschura$^{a)}$, E.~J.~Weniger$^{b),c)}$,  G.~Soff$^{a)}$
\end{center}

\vspace{0.2cm}

\begin{center}
$^{a)}${\it Institut f\"ur Theoretische Physik,}\\
{\it Technical University of Dresden, 01062 Dresden, Germany} \\[1ex]
$^{b)}${\it Institut f\"{u}r Physikalische und Theoretische Chemie,}\\
{\it Universit\"{a}t Regensburg, 93040 Regensburg, Germany}\\[1ex]
$^{c)}${\it Max-Planck-Institut f\"{u}r Physik komplexer Systeme,}\\
{\it N\"{o}thnitzer Stra\ss e 38--40, 01087 Dresden, Germany}
\end{center}

\vspace{1.3cm}

\begin{center}
\begin{minipage}{10.5cm}
{\underline{Abstract}} The improvement of resummation algorithms for
divergent perturbative expansions in quantum field theory 
by asymptotic information about
perturbative coefficients is investigated. 
Various asymptotically optimized
resummation prescriptions are
considered. The improvement of perturbative predictions beyond the
reexpansion of rational approximants is discussed. 
\end{minipage}
\end{center}

\vspace{1.3cm}

Asymptotic problems and properties;\\
Quantum Electrodynamics -- Specific Calculations\\
\newpage

\typeout{::Section 1}
%
%
\section{INTRODUCTION}
\label{Sec:intro}

There is overwhelming evidence that perturbation series in quantum field
theory and related disciplines diverge (see for
example~\cite{LeGuZiJu1990} and references therein). Consequently,
resummation techniques, which allow to associate a finite value to a
divergent series, are needed if divergent perturbation series are to be
used for numerical purposes. The best known and most often applied
resummation techniques are Pad\'e approximants \cite{BaGr1996} and the
Borel method~\cite{Bor1899,Bor1928}. Recently, another summation method
-- the so-called delta transformation (Eq.\ (8.4-4) of \cite{We1989}) --
has gained some prominence as a summation method in various domains of
physics~\cite{Delta1,Delta2,Delta3,JeMoSoWe1999,WeCiVi1993,We1997,
JeBeWeSo1999}. There is evidence~\cite{Delta2,WeCiVi1993} that the delta
transformation is able to sum divergent series whose coefficients $c_n$
grow essentially like $n!$, $(2n)!$ and even $(3 n)!$. This is not
achievable by employing Pad\'e approximants~\cite{GrafGrec1978}.

In the case of both Pad\'e approximants and the delta transformation,
only the numerical values of a finite number of partial sums of the
divergent input series are needed as input data. Within the framework of
the Borel method, it is also necessary to know the large order
asymptotics of the coefficients of the divergent power series. 
Consequently, the Borel method is slightly less general than the
other two methods.

Rational approximations to (divergent) power series have an interesting
feature: It is possible to extrapolate higher-order coefficients that
were not used for the construction of the rational approximants. This
feature, which was apparently first observed by Gilewicz \cite{Gil1973},
has so far been used quite extensively in the case of Pad\'{e}
approximants for divergent perturbation expansions from quantum field
theory and related disciplines.  We refer to the investigations by
Elias, Steele and Chishtie {\em et al.}~\cite{ElStChMiSp1998,StEl1998,
ChElMiSt1999,ChElSt1999a,ChElSt1999b,ChElSt2000,ElChSt2000}, by
Samuel, Gardi, Karliner, Ellis {\em et
al.}~\cite{SaLi1991,SaLiSt1993,Sa1994,SaLiSt1994,
SaLi1994a,SaLi1994b,SaElKa1995,Sa1995a,Sa1995b,
SaLiSt1995,ElGaKaSa1996,ElGaKaSa1996a,ElGaKaSa1996b,
Ga1997,ElKaSa1997,SaAbYu1997,JaJoSa1997,Karliner1998,
ElJaJoKaSa1998,BuAbSaStMa1996} and by Cveti\v{c} {\em et
al.}~\cite{CvKo1998,Cv1998a,Cv1998b,CvYu2000}. An analogous approach
also works in the case of the delta
transformation~\cite{We1997,JeBeWeSo1999}. Recursive techniques for the
Pad\'{e} prediction of unknown series coefficients were developed
recently~\cite{We2000}.

It is the intention of this article to discuss -- augmenting previous
investigations~\cite{JeBeWeSo1999} -- how additional information on the
large-order asymptotics of the perturbative coefficients can be
incorporated effectively into resummation and prediction schemes. Our
emphasis will be on the Borel-Pad\'e method which was introduced by
Graffi, Grecchi, and Simon~\cite{GraGreSim1970}. This is a variant of
the Borel method that uses Pad\'e approximants for performing the
analytic continuation of the Borel transformed series to a neighborhood
of the positive real semiaxis. Further details on the Borel-Pad\'e
method as well as on other resummation methods can be found in
Section~\ref{Sec:ResumMeth}.

We would like to mention here the investigations (based on the Borel
method) by Fischer~\cite{Fi1997polo,Fi1997,Fi1999} and by Caprini and
Fischer~\cite{CaFi1999,CaFi2000} regarding the resummation of divergent
perturbative expansions in quantum field theory. In~\cite{CaFi1999},
Caprini and Fischer use asymptotic information (location of the first IR
and UV renormalon poles) for the construction of conformal mappings in
the Borel plane. Here, we describe alternative ways in which the
asymptotics of the coefficients of the perturbative expansion can be
utilized to enhance the effectiveness of resummation procedures and
perturbative predictions. These improvements, which provide an
alternative to the methods presented in~\cite{CaFi1999}, are not
necessarily restricted to the first IR and UV renormalons, but can take
advantage of the location of all known poles in the Borel plane.
Regarding asymptotic properties of perturbative coefficients in
quantum field theory we also refer to~\cite{West1991,West1999}.

Cveti\v{c} and Yu in~\cite{CvYu2000} consider the resummation of the
\emph{real} part of the (one-loop) QED effective action (or vacuum-to
vacuum amplitude) in the presence of background electric and magnetic
fields, of which the exact nonperturbative answer is known
\cite{Sc1951}. In the presence of an electric field, the QED effective
action acquires an imaginary part proportional to the pair production
amplitude for real electron-positron (or lepton-antilepton) pairs. The
real part of the effective action does not constitute the full physical
solution. The full, complex-valued answer requires an integration in
the complex plane, for example along the special contour introduced
in~\cite{Je2000}. Only
in the case of the pure magnetic field, which we consider in this
article, the QED effective action is entirely real.  For this particular
example, the delta transformation employed in~\cite{JeBeWeSo1999} or the
Borel-Pad\'{e} Cauchy principal value method used in~\cite{CvYu2000}
provide the full, i.e.~entirely real, not complex physical solution.  In
view of the above, we again consider here only the resummation of the
pure magnetic field. The electric field or the electric field combined
with a magnetic field should be treated along the ideas introduced
in~\cite{Je2000}.

We briefly discuss the relation of resummation and perturbative
predictions: There is evidence (see for example
\cite{JeBeWeSo1999,Je2000} and references therein) that the
\emph{resummation} of divergent series is ambiguous, especially if 
these series do not fulfill a Carleman condition (see \cite{Car1926} or
Theorems XII.17 and XII.19 in~\cite{ReSi1978}). However, the
\emph{prediction} of perturbative coefficients apparently does not 
suffer from such ambiguities. Perturbative predictions should be
possible~\cite{JeBeWeSo1999} even in those cases where the perturbation
series is evaluated ``on the cut'' in the complex plane. For the
prediction of unknown perturbation series coefficients, the nonanalytic
contributions, which are responsible for the ambiguities, are
irrelevant. Only the analytic part of the function, which is represented
by a divergent power series, matters.  The terms of an otherwise more
problematic divergent nonalternating series can be predicted just as
well as the terms of an alternating series.  That is to say, the
resummation of a divergent series is not necessarily unique (see the
different integration contours in~\cite{Je2000}), but the perturbative
coefficients, which can be extrapolated and predicted via rational
approximants, are uniquely determined even though the complex
integration along the different contours in~\cite{Je2000} leads to
different results for the nonperturbative, nonanalytic contributions.

This paper is organized as follows. In Section~\ref{Sec:ResumMeth} we
present a short account of various resummation methods for divergent
series. The exploitation of additional available asymptotic information
in resummation algorithms is discussed in Section~\ref{Sec:AsyOptRes}.
Asymptotically optimized perturbative predictions (i.e., predictions of
higher-order unknown perturbative coefficients) are discussed in
Sections~\ref{Sec:AsyOptPre} and~\ref{Sec:PerSerAno} for various example
cases. We conclude with a discussion of the results in
Section~\ref{Sec:Conclu}.

\typeout{::Section 2}
%
%
\section{A REVIEW OF RESUMMATION METHODS}
\label{Sec:ResumMeth}

In this Section, we provide a condensed description of the
resummation methods under consideration in this article. Let
\begin{equation}
\label{PowSer_f}
f (z) \; \sim \; \sum_{\nu=0}^{\infty} \, \gamma_{\nu} \, z^{\nu}	
\end{equation}
be a (formal) power series for some function $f$. Then, we define the
$(\kappa, \lambda)$-generalized Borel integral transform according
to
\begin{eqnarray}
\lefteqn{} \nonumber \\
f (z) & = & \int_{0}^{\infty} \, t^{\lambda - 1} \, 
\mathcal{B}^{(\kappa, \lambda)} (f; z t^{\kappa}) \,
\exp \bigl(- t \bigr) \, 
\mathrm{d} t
\label{BorIntTran_1} 
\\
& = & z^{-\lambda/\kappa} \, \int_{0}^{\infty} \, s^{\lambda - 1} \, 
\mathcal{B}^{(\kappa, \lambda)} (f; s^{\kappa}) \,
\exp \bigl(- s/z^{1/k} \bigr) \,
\mathrm{d} s \, .
\label{BorIntTran_2}
\end{eqnarray}
Here,
\begin{equation}
\label{BorTranSer} 
\mathcal{B}^{(\kappa, \lambda)} (f; z) \; = \; \sum_{\nu=0}^{\infty} \, 
\frac{\gamma_{\nu}}{\Gamma (\kappa \nu + \lambda)} \, z^{\nu}
\end{equation}
is the $(\kappa, \lambda)$-generalized Borel transformed series of the
power series (\ref{PowSer_f}) for $f (z)$. For $\kappa = \lambda = 1$,
we recover the usual formulas for the Borel transformation:
\begin{eqnarray}
f (z) & = & \int_{0}^{\infty} \, 
\mathcal{B} (f; z t) \, \exp \bigl(- t \bigr) \, \mathrm{d} t
\\
& = & \frac{1}{z} \, \int_{0}^{\infty} \,
\mathcal{B} (f; s) \,\exp \bigl(- s/z \bigr) \, \mathrm{d} s \, ,
\\
\mathcal{B} (f; z) & = & \mathcal{B}^{(1, 1)} (f; z) 
\; = \; \sum_{\nu=0}^{\infty} \, 
\frac{\gamma_{\nu}}{{\nu}!} \, z^{\nu} \, .
\end{eqnarray}

There exists an extensive literature on the Borel method in general and
on physical applications in special. Any attempt to provide something
resembling a reasonably complete bibliography would clearly be beyond
the scope of this article. Let us just mention that recent monographs on
the Borel method and related topic were published by Shawyer and Watson
\cite{ShaWat1994} and Sternin and Shatalov \cite{SteSha1996}.

Let us assume that the coefficients $\gamma_n$ of the power series
(\ref{PowSer_f}) possess the following large order asymptotics,
\begin{equation}
\gamma_n \; \sim \; A \, \Gamma (\kappa n + \lambda) B^n \, ,
\qquad n \to \infty \, , 
\label{AsyCf}
\end{equation}
where $A$, $B$, $\kappa$, and $\lambda$ are suitable constants. We say
that a $(\kappa', \lambda')$-generalized Borel method for some power
series (\ref{PowSer_f}) is \emph{asymptotically optimized} if the
parameters $\kappa'$ and $\lambda'$ agree with the parameters $\kappa$
and $\lambda$ in the large-order asymptotics (\ref{AsyCf}) for the
series coefficient $\gamma_n$. Thus, the leading (hyper)factorial growth
of $\gamma_n$ is exactly canceled out in this case. If we know in
addition the parameter $B$ in (\ref{AsyCf}), then we can immediately
deduce that the asymptotically optimized Borel transformed series
(\ref{BorTranSer}) possesses a pole at $z = 1/B$.

The most difficult computational problem, which normally occurs in the
context of a Borel summation process, is the construction of an analytic
continuation for the Borel transformed series (\ref{BorTranSer}). If the
coefficients $\gamma_n$ of the power series (\ref{PowSer_f}) satisfy
(\ref{AsyCf}), then the $(\kappa, \lambda)$-generalized Borel
transformed series (\ref{BorTranSer}) has a \emph{nonzero} but
\emph{finite} radius of convergence. In order to be able to do the
integration, we now need an analytic continuation which extends
$\mathcal{B}^{(\kappa, \lambda)} (f; z)$ from the interior of its circle
of convergence to a neighborhood which contains the whole positive
semiaxis. In this article, we emphasize the Borel-Pad\'e method which
was introduced by Graffi, Grecchi, and Simon \cite{GraGreSim1970} and
which accomplishes the analytic continuation by converting the partial
sums of the Borel transformed series (\ref{BorTranSer}) to Pad\'e
approximants. It may happen that the Pad\'{e} approximants thus
constructed exhibit poles along the positive real axis. In such a case,
the function $f$ is -- strictly speaking -- not Borel-Pad\'{e}
summable. However, by further generalizing the integrals
(\ref{BorIntTran_1}) and (\ref{BorIntTran_2}) -- either via a
principal-value prescription~\cite{Raczka1991,Pi1999}, or by employing
conformal mappings~\cite{CaFi1999,CaFi2000}, or via the special
integration contours used in~\cite{Je2000} -- it may nevertheless be
possible to associate a finite value to the divergent power series
(\ref{PowSer_f}) for $f$.

In the case of Pad\'e approximants we use the notation and conventions
of the monograph by Baker and Graves-Morris \cite{BaGr1996}. Thus, a
Pad\'{e} approximant $[ l / m ]_f (z)$ to $f (z)$ corresponds to the
ratio of two polynomials $P_l (z)$ and $Q_m (z)$, which are of degrees
$l$ and $m$, respectively, in $z$:
\begin{equation}
[ l / m ]_f (z) \;=\; \frac{P_l (z)}{Q_m (z)} \; = \;
\frac{p_0 + p_1 \, z + \ldots + p_{l} \,  z^{l}}
{1 + q_1 \, z + \ldots + q_{m} \, z^{m}} \, .
\label{DefPade}
\end{equation}
The polynomials $P_l (z)$ and $Q_m(z)$ are constructed so that the
Taylor expansion of the Pad\'e approximation agrees with the original
input series (\ref{PowSer_f}) up to terms of order $l+m$ in $z$,
\begin{equation}
\label{O_est_PA} 
f (z) \, - \, [ l / m ]_{f} (z) \; = \; 
{\rm O} \bigl(z^{l+m+1} \bigr)\, , \qquad z \to 0 \, .
\end{equation}
This asymptotic error estimate leads to a system of linear equations by
means of which the coefficients $p_0$, $p_1$, \ldots, $p_l$ and $q_1$,
$q_2$, \ldots, $q_m$ in (\ref{DefPade}) can be computed. However, there
are also several algorithms which permit a recursive computation of
Pad\'e approximants. A discussion of the merits and weaknesses of the
various computational approaches can for instance be found in Section
II.3 of the book by Cuyt and Wuytack \cite{CuyWuy1987}.

An example of such a recursive algorithm is provided by Wynn's
epsilon algorithm \cite{Wy1956}:
\begin{eqnarray}
\label{eps_al}
\epsilon_{-1}^{(n)} & \; = \; & 0 \, ,
\qquad \epsilon_0^{(n)} \, = \, s_n \, ,
\qquad  n \in \bbox{N}_0 \, , \\
\epsilon_{k+1}^{(n)} & \; = \; & \epsilon_{k-1}^{(n+1)} \, + \,
1 / [\epsilon_{k}^{(n+1)} - \epsilon_{k}^{(n)} ] \, ,
\qquad k, n \in \bbox{N}_0 \, .
\end{eqnarray}
\noindent
Wynn~\cite{Wy1956} showed that if the input data $s_n$ for the epsilon
algorithm are the partial sums of the (formal) power series
(\ref{PowSer_f}) for some function $f (z)$ according to
\begin{equation}
s_n = f_n (z) \; = \; \sum_{\nu=0}^{n} \gamma_{\nu} z^{\nu} \, ,
\label{PowSerPS}
\end{equation}
then the elements $\epsilon_{2 k}^{(n)}$ with {\em even\/} subscripts
are Pad\'{e} approximants to $f$ according to
\begin{equation}
\epsilon_{2 k}^{(n)} \; = \; [ n + k / k ]_f(z) \, .
\label{Eps_Pade}
\end{equation}
In contrast, the elements $\epsilon_{2 k + 1}^{(n)}$ with {\em odd\/}
subscripts are only auxiliary quantities which diverge if the whole
process converges.

If one tries to sum a divergent power series or to accelerate the
convergence of a slowly convergent series by converting its partial sums
to Pad\'e approximants, it is usually a good idea to use either {\em
diagonal} Pad\'{e} approximants, whose numerator and denominator
polynomials have equal degrees, or -- if this is not possible -- to use
Pad\'{e} approximants with numerator and denominator polynomials whose
degrees differ as little as possible. If we use the epsilon algorithm
for the computation of the Pad\'e approximants, then Eq.\
(\ref{Eps_Pade}) implies that we then obtain the following staircase
sequence in the Pad\'e table (see Eq.\ (4.3-7) of \cite{We1989}):
\begin{equation}
[0/0], [1/0], [1/1], \ldots \, , [\nu / \nu], [\nu + 1/ \nu], 
[\nu +1/ \nu +1], \ldots \, .  
\end{equation}
This staircase sequence exploits the available information optimally if
the partial sums $f_m (z)$ with $m \ge 0$ are computed successively and
if after the computation of each new partial sum the element of the
epsilon table with the highest possible {\em even\/} transformation
order is computed. With the help of the notation $\trunc{x}$ for the
integral part of $x$, this staircase sequence can be written compactly
as follows:
\begin{equation}
\epsilon_{2 \trunc{n/2}}^{(n - 2 \trunc{n/2})} \; = \;
\bigl[ n - \trunc{n/2} / \trunc{n/2} \bigr]_f (z) \, ,
\qquad n = 0, 1, 2, \ldots \, .  
\end{equation}

The asymptotic error estimate (\ref{O_est_PA}) implies that all series
coefficients, which are employed for the computation of the Pad\'e
approximant $[ l / m ]_{f} (z)$, are recovered by a Taylor
expansion. Consequently, the higher order derivatives of the Pad\'e
approximant provide predictions for ``unknown'' series coefficients,
i.e.\ to those series coefficients that were not used for computation of
$[ l / m ]_{f} (z)$.

There is an enormous amount of literature on Pad\'{e} approximants in
general as well as on their application in theoretical physics. Let us
just mention that the popularity of Pad\'{e} approximants in theoretical
physics can be traced back to a review by Baker~\cite{Ba1965}, that the
monograph by Baker and Graves-Morris~\cite{BaGr1996} is the currently
most complete source of information on Pad\'{e} approximants, and that
an account of the historical development of Pad\'{e} approximants and
related topics is given in a monograph by Brezinski~\cite{Bre1991}.

The intense research on Pad\'e approximants during the last decades of
course also showed that Pad\'e approximants suffer -- like all other
numerical techniques -- from certain limitations and weaknesses. For
example, Pad\'e approximants are in principle limited to convergent or
divergent power series, but cannot help in the case of many other types
of slowly convergent or divergent sequences. Moreover, Pad\'e
approximants are either not useful or cannot be applied at all in the
case of power series whose coefficients $\gamma_n$ grow like $(2 n)!$ or
even $(3 n)!$~\cite{GrafGrec1978}. Consequently, the intense research on
Pad\'e approximants also stimulated research on related techniques, the
so-called sequence transformations.

Let us assume that $\{ s_n \}_{n=0}^{\infty}$ is a sequence, whose
elements may for instance be the partial sums of an infinite series
according to $s_n = \sum_{k=0}^{n} a_k$. A \emph{sequence
transformation} is a rule which maps a sequence $\{ s_n
\}_{n=0}^{\infty}$ to a new sequence $\{ s'_n \}_{n=0}^{\infty}$ with
hopefully better numerical properties. In this terminology, Pad\'{e}
approximants are just a special class of sequence transformations since
they transform the partial sums of a (formal) power series to a doubly
indexed sequence of rational approximants.

If $\{ s_n \}_{n=0}^{\infty}$ either converge to some limit $s$ as $n
\to \infty$ or can be summed to the generalized limit $s$ in the case of
divergence, then a sequence element $s_n$ can for all $n \ge 0$ be
partitioned into the (generalized) limit $s$ and a remainder $r_n$
according to
\begin{equation}
s_n \; = \; s \, + \, r_n \, .
\end{equation}
Normally, a sequence transformation will not be able to determine the
(generalized) limit $s$ of $\{ s_n \}_{n=0}^{\infty}$
\emph{exactly}. Thus, the elements of the transformed sequence $\{ s'_n
\}_{n=0}^{\infty}$ can also be partitioned into the (generalized) limit
$s$ and a transformed remainder $r'_n$ according to
\begin{equation}
s'_n \; = \; s \, + \, r'_n \, ,
\end{equation}
and the transformed remainders will in general be different from zero
for all finite values of $n$.

In the literature on convergence acceleration it is said that a sequence
transformation \emph{accelerates convergence} if the transformed
remainders $\{ r'_n \}_{n=0}^{\infty}$ vanish more rapidly than the
original remainders $\{ r_n \}_{n=0}^{\infty}$ according to
\begin{equation}
\lim_{n \to \infty} \, \frac{r'_n}{r_n} \; = \; 
\lim_{n \to \infty} \, \frac{s'_n - s}{s_n - s} \; = \; 0 \, ,
\end{equation}
and a divergent sequences, whose remainders $r_n$ do not vanish as $n
\to \infty$, is summed to its generalized limit $s$ if the transformed
remainders $r'_n$ vanish as $n \to \infty$. 

Thus, a sequence transformation essentially tries to eliminate the
remainders $r_n$ from the sequence elements $s_n$ as effectively as
possible. Since, however, an in principle unlimited variety of different
remainders can occur, it is necessary to make some assumptions -- either
explicitly or implicitly -- which provide the basis for the construction
of a sequence transformation. A detailed discussion of the construction
of sequence transformations as well as many examples can be found in the
book by Brezinski and Redivo Zaglia~\cite{BreRZa1991} or
in~\cite{We1989}.

Normally, the assumptions being made are incorporated into the
transformation scheme via model sequences, whose remainders possess a
particular simple structure and can be expressed by a finite number of
terms:
\begin{equation}
\tilde{s}_n \; = \; \tilde{s} \, + \, 
\sum_{k=0}^{k-1} \, \tilde{c}_j \, \varphi_j (n) \, .
\label{GenModSeqPhi}
\end{equation}
Here, the $\tilde{c}_j$ are unspecified coefficients, and the $\varphi_j
(n)$ are by assumption known functions of $n$. 

The elements of this model sequence contain $k+1$ unknown, the
(generalized) limit $\tilde{s}$ and the $k$ coefficients $\tilde{c}_j$
with $0 \le j \le k-1$. Since all unknowns occur \emph{linearly}, it is
possible to construct a sequence transformation $\mathcal{T}$ -- if
necessary via Cramer's rule -- which is \emph{exact} for the elements of
this model sequence according to
\begin{equation}
\mathcal{T} \; = \; \mathcal{T} 
\left( \tilde{s}_{n}, \tilde{s}_{n+1}, \ldots, \tilde{s}_{n+k}, \right) 
\; = \; \tilde{s} \, ,
\end{equation}
if applied to the numerical values of $k + 1$ sequence elements
$\tilde{s}_{n}, \tilde{s}_{n+1}, \ldots, \tilde{s}_{n+k}$.

Of course, simple model sequences of that kind normally do not occur in
practical problems. However, their elements provide at least for
sufficiently large values of $k$ reasonably accurate approximations to
the elements of the more realistic sequence
\begin{equation}
s_n \; = \; s \, + \, \sum_{j=0}^{\infty} \, c_j \, \varphi_j (n) \, .
\label{SeqPhi}
\end{equation}
If we now apply this sequence transformation $\mathcal{T}$ to the
numerical values of $k+1$ elements of the sequence (\ref{SeqPhi}), then
we have no reason to assume that $\mathcal{T}$ might produce its exact
(generalized) limit $s$. However, a more detailed mathematical analysis
of the transformation process normally reveals that $\mathcal{T}$
eliminates the first $k$ terms $c_j \varphi_j (n)$ with $0 \le j \le
k-1$. Thus, the transformed remainder $r'_n$ starts with $\varphi_k (n)$
instead of $\varphi_0 (n)$, which for sufficiently large values of $k$
normally constitutes a significant improvement.

Most sequence transformations can be constructed on the basis of model
sequences of the type of Eq.\ (\ref{GenModSeqPhi}). For example, Wynn
\cite{Wy1956} could show that his epsilon algorithm is exact for model
sequences whose remainders can be expressed as a linear combination of
exponential terms according to
\begin{equation}
\tilde{s}_n \; = \; \tilde{s} \, + \, 
\sum_{j=0}^{k-1} \, \tilde{c}_j \, \lambda_{j}^{n} \, .
\label{ModSeqEps}
\end{equation}
Concerning the $\lambda_j$ it is only assumed that they are different
from zero and one and ordered according to magnitude, i.e., $\lambda_j
\ne 0, 1$ and $\vert \lambda_0 \vert > \vert \lambda_1 \vert > \vert 
\lambda_{k-1} \vert > 0$. Thus, if the numerical values of $2k+1$ 
elements $\tilde{s}_{n}$, $\tilde{s}_{n+1}$, \ldots, $\tilde{s}_{n+2k}$
of this model sequence are available, then the epsilon algorithm is
exact according to
\begin{equation}
\epsilon_{2k}^{(n)} \; = \; \tilde{s} \, .
\end{equation}
Moreover, Wynn constructed in Theorems 16 and 17 of~\cite{Wy1966}
asymptotic expansions ($n \to \infty$) for the transformed 
remainders $r'_n$ created by the
application of $\epsilon_{2k}^{(n)}$ to the elements of the sequence
\begin{equation}
s_n \; = \; s \, + \, \sum_{j=0}^{\infty} \, c_j \, \lambda_{j}^{n} \, ,
\label{GenModSeqEps}
\end{equation}
which is an obvious generalization of the model sequence
(\ref{ModSeqEps}). He showed that the transformed remainders $r'_n$ are
proportional to $\lambda_{k}^{n}$ which corresponds to an elimination of
the first $k$ exponential terms $c_j \lambda_{j}^{n}$ on the right-hand
side of (\ref{GenModSeqEps}). Since the $\lambda_j$ are by assumption
ordered in magnitude, this constitutes a significant
achievement. Consequently, Wynn's epsilon algorithm is
\emph{asymptotically optimal} for sequences of the type of
(\ref{GenModSeqEps}). This means that no other sequence transformation,
which also uses only the numerical values of the elements of the
sequence (\ref{GenModSeqEps}) as input data, can produce a better
asymptotic ($n \to \infty$) truncation error.

Levin \cite{Lev1973}
introduced a class of sequence transformations which are exact for model 
sequences of the following type:
\begin{equation}
\tilde{s}_n \; = \; \tilde{s} \, + \, \omega_n \, z_n \, .
\label{LevTypeModSeq}
\end{equation}
Here, $\omega_n$ is an estimate for the truncation error $\tilde{r}_n$,
and $z_n$ is a correction term. Levin \cite{Lev1973} assumed that $z_n$
can be expressed as a truncated power series in $1/(n+\zeta)$ where
$\zeta$ is a positive shift parameter. In Sections 7 -- 9 of
\cite{We1989}, several other sequence transformations were
constructed which are also exact for the model sequence
(\ref{LevTypeModSeq}) but make different assumptions about the
correction terms $z_n$. 
The remainder estimates $\omega_n$ introduce additional
degrees of freedom in the construction of the sequence
transformation as compared to Pad\'{e} approximants.
One may draw an analogy between sequence
transformations and Pad\'{e} approximants on the 
one hand and the Gaussian integration and
the Simpson rule on the other hand; 
the variable integration nodes and weight factors
of the Gaussian integration
yield additional degrees of freedom which may be used in order
to construct a potentially much more powerful algorithm for
numerical integration.        

In the following text, we will concentrate on
sequence transformations which assume that $z_n$ can be expressed as a
truncated factorial series (Section 8 of \cite{We1989}):
\begin{equation}
z_n \; = \; \sum_{j=0}^{k-1} \, \tilde{c}_j / (\zeta+n)_j \, .
\label{ModSeqFactSer}
\end{equation}
Here, $(n+\zeta)_j = \Gamma(n+\zeta+j)/\Gamma(n+\zeta)$ is a Pochhammer
symbol, and $\zeta$ is a positive shift parameter. The assumption
(\ref{ModSeqFactSer}) implies that the sequence transformations derived
in this way are particularly well suited for sequences satisfying
\begin{equation}
s_n \; = \; s \, + \, \omega_n \, 
\sum_{j=0}^{\infty} \, c_j / (\zeta+n)_j \, .
\label{SeqFacSer}
\end{equation}
It is {\em a priori} not obvious that the ratio $[s_n - s]/\omega_n =
r_n/\omega_n$ can be expressed as a factorial series.
Nevertheless, this assumption leads to powerful sequence transformations
which are apparently particularly well suited for the summation of
factorially divergent 
series~\cite{Delta1,Delta2,Delta3,WeCiVi1993,We1997,JeBeWeSo1999}. 
Let $s_n = \sum_{k=0}^{n} a_k$ be the partial sums of an infinite
series. If the $a_k$ strictly alternate in sign and decrease
monotonously in magnitude, then the best simple estimate for the
truncation error $r_n = - \sum_{k=n+1}^{\infty} a_k$ is the first term
$a_{n+1}$ neglected in the partial sum $s_n$. Moreover, the first term
neglected is also the best simple remainder estimate for many
factorially divergent alternating series (see for example Theorem 13-2
of \cite{We1989}). The mathematical structure of a factorially
divergent series is expected for the perturbative expansions in quantum
field theory. Further arguments supporting the general applicability
of the delta transformation to the series
of the mathematical structure as expected for quantum field theory
will be discussed in~\cite{JeBeMHMoWeSo1999}.

If we now combine the assumption that $z_n$ should be a truncated
factorial series according to (\ref{ModSeqFactSer}) with the remainder
estimate 
\begin{equation}
\label{omegaN}
\omega_n \; = \; \Delta s_n \; = \; a_{n+1} \, ,
\end{equation}
which was introduced by Smith and Ford \cite{SmiFo1979}, we obtain the
delta transformation which is defined by the following ratio of finite
sums (Eq.\ (8.4-4) of \cite{We1989}):
\begin{eqnarray}
{\delta}_{k}^{(n)} (\zeta, s_n) & = & \frac
{\Delta^k \left[ (\zeta+n)_{k-1} s_n /\Delta s_n \right]}
{\Delta^k \left[ (\zeta+n)_{k-1}/\Delta s_n \right]}
\nonumber
\\
& = & \frac
{\displaystyle
\sum_{j=0}^{k} \; ( - 1)^{j} \; {{k} \choose {j}} \;
\frac {(\zeta + n +j )_{k-1}} {(\zeta + n + k )_{k-1}} \;
\frac {s_{n+j}} {\Delta s_{n+j}} }
{\displaystyle
\sum_{j=0}^{k} \; ( - 1)^{j} \; {{k} \choose {j}} \;
\frac {(\zeta + n +j )_{k-1}} {(\zeta + n + k )_{k-1}} \;
\frac {1} {\Delta s_{n+j}} }
\; .    
\label{deltaTr}
\end{eqnarray}
Here, the same notation as in~\cite{We1989} is used. Thus, $\Delta$
stands for the difference operator defined by $\Delta g (n) = g (n+1) -
g (n)$, $(a)_n = \Gamma(a+n)/\Gamma(a)$ is a Pochhammer symbol, $k$ and
$n$ are nonnegative integers, and $\zeta$ is a shift parameter which has
to be positive to allow $n = 0$ in Eq. (\ref{deltaTr}). The most obvious
choice, which is always used in this article, is $\zeta = 1$.

In Section 8.3 of \cite{We1989}, a simple recursive scheme is described
which permits -- depending upon the initial values -- the recursive
calculation of either the numerator or the denominator sum of
${\delta}_{k}^{(n)} (\zeta, s_n)$.

In the context of quantum field theory and related disciplines, the
delta transformation (\ref{deltaTr}) may be used for the summation
of divergent perturbation expansions which are power series in some
coupling constant. Thus, if we replace the input data $s_n$ in
(\ref{deltaTr}) by the partial sums $f_n (z) = \sum_{\nu=0}^{n}
\gamma_{\nu} z^{\nu}$ of the (formal) power series (\ref{PowSer_f}) for
$f (z)$, we obtain a rational expression, whose numerator and
denominator polynomials are of degrees $k+n$ and $k$, respectively, in
$z$:
\begin{equation}
{\delta}_k^{(n)} \bigl(\zeta, f_n (z) \bigr) \; = \;
\frac
{\displaystyle
\sum_{j=0}^{k} \; ( - 1)^{j} \; {{k} \choose {j}} \;
\frac
{(\zeta + n +j )_{k-1}} {(\zeta + n + k )_{k-1}} \;
\frac {z^{k - j} f_{n+j} (z)} {\gamma_{n+j+1}}}
{\displaystyle
\sum_{j=0}^{k} \; ( - 1)^{j} \; {{k} \choose {j}} \;
\frac
{(\zeta + n +j )_{k-1}} {(\zeta + n + k )_{k-1}} \;
\frac {z^{k - j}} {\gamma_{n+j+1}}} \, .
\label{deltaPS} 
\end{equation}
If the coefficients $\gamma_n$ of the power series for $f (z)$ are all
different from zero, the rational function (\ref{deltaPS}) satisfies the
asymptotic error estimate (Eq.\ (4.29) of~\cite{WeCiVi1993})
\begin{equation}
\label{O_est_delta}
f (z) \, - \, {\delta}_k^{(n)} \bigl(\zeta, f_n (z) \bigr)
\; = \; O (z^{k + n + 2}) \, , \quad z \to 0 \, . 
\end{equation}
This estimate, which is formally very similar to the analogous estimate
(\ref{O_est_PA}) for Pad\'{e} approximants, implies that all terms of
the formal power series, which are used for construction of the rational
approximant ${\delta}_k^{(n)} \bigl(\zeta, f_n (z) \bigr)$, are
reproduced exactly by a Taylor expansion around $z = 0$. Moreover, the
higher order derivatives provide just like in the Pad\'{e} case
predictions for those coefficients $\gamma_{n+k+2}$, $\gamma_{n+k+3}$,
\ldots, that were not used for the construction of the rational
function. 

As already discussed, the power of the delta transformation or of other
Levin-type transformations results from the fact that an explicit
estimate for the truncation error is incorporated into the
transformation scheme. The truncation error estimate used by the delta
transformation is the first term $\gamma_{n+1} z^{n+1}$ neglected in the
partial sum $f_n (z) = \sum_{\nu=0}^{n} \gamma_{\nu} z^{\nu}$. 
Consequently, for a proper application of the delta transformation
\emph{all} coefficients $\gamma_n$ of the power
series for $f$ with $n \ge 1$ have to be different from zero because
otherwise the estimate for the truncation error makes no sense. This
restriction also follows directly from the ratio (\ref{deltaPS}), where
undefined expressions occur if coefficients $\gamma_n$ with $n \ge 1$
are zero (cf.~Eq.~(11) in~\cite{CvYu2000} which
entails divisions by zero).       

\typeout{::Section 3}
%
%
\section{ASYMPTOTICALLY OPTIMIZED RESUMMATION}
\label{Sec:AsyOptRes}

The problem of the resummation of divergent perturbative expansions in
quantum field theory and related disciplines has been discussed in a
number of recent publications, for example in~\cite{CvYu2000,Je2000,
CvYu1999,CaFi1999,Raczka1991,Pi1999}. We investigate here asymptotically
optimized resummation methods, i.e.~methods which utilize information about
large-order asymptotics of perturbative coefficients with the intention
of enhancing the rate of convergence of the resummation
algorithm. 

We discuss here possible improvements of the Borel-Pad\'{e} method on
the basis of potentially available information about the large-order
asymptotics of perturbative coefficients. We concentrate on the
particular model example discussed recently by 
Dunne and Hall~\cite{DuHa1999}, by Cveti\v{c} and Yu in~\cite{CvYu2000}
and by ourselves in~\cite{JeBeWeSo1999}.
We discuss the QED effective action
$S_{\mathrm{B}}$ in the presence of a constant background magnetic
field. The exact nonperturbative result for $S_{\mathrm{B}}$ can be
expressed as a proper-time integral:
\begin{equation}
\label{SRB}
S_{\mathrm{B}} = - \frac{e^2 \, B^2}{8 \pi^2} \!
\int\limits_0^\infty \frac{{\rm d} s}{s^2} \!
\left\{\coth s - \frac{1}{s} - \frac{s}{3} \right\}
\exp\!\left(-\frac{m_{\rm e}^2}{e \, B} s\right)\,.
\end{equation}
Here, $B$ is the magnetic field strength, and $m_{\mathrm{e}}$ and $e$
are the mass and the charge of the electron, respectively
(this result is given for example in Eq.~(4-123) in~\cite{ItZu1980}). 
 
The integral representation (\ref{SRB}) for $S_{\mathrm{B}}$ can be
expressed as a strictly alternating perturbation series in the effective
coupling coupling $g_{\mathrm{B}} = e^2 B^2 / m_{\mathrm{e}}^4$:
\begin{equation}
\label{Bperser}
S_{\mathrm{B}} = - \, \frac{2 e^2 B^2 }{\pi^2} \, g_{\mathrm{B}} \,
\sum_{n=0}^\infty \; c_n \; g_{\mathrm{B}}^n \, ,
\end{equation}
where
\begin{equation}
\label{cn}                    
c_n = \frac{(-1)^{n + 1} \, 4^n \,
\left| {\cal B}_{2n+4} \right| }{(2n+4)(2n+3)(2n+2)} \, .
\end{equation}
Here, ${\cal B}_{2n+4}$ is a Bernoulli number. Thus, the
perturbation expansion (\ref{Bperser}) has the remarkable feature that an
unlimited number of series coefficients $c_n$ are known analytically. 
Consequently, this series is particularly well suited as a model
system for studying resummation methods.

Next, we utilize the fact that a Bernoulli number with even index can be
expressed by a Riemann zeta function according to (see Section~23.2.~on
p.~807 of~\cite{AbSt1972})
\begin{equation}
\left| {\cal B}_{2n} \right| \; = \;
\frac{2\,(2n)!}{(2\pi)^{2n}} \, \zeta (2n) \, .
\end{equation}
Inserting this into (\ref{Bperser}) and (\ref{cn}) yields
\begin{eqnarray}
\label{S_B_1}
S_{\mathrm{B}} &=& - \frac{e^2 B^2}{8 \pi^2} \, g_{\rm B} \,
\sum_{n=0}^{\infty} \, (-1)^n \,
(2n+1)! \, \frac{2\,\zeta(2n+4)}{\pi^{2n+4}} \, g_{\rm B}^n \\
\label{defSBprime1}
&=& - \frac{e^2 B^2}{8 \pi^2} \, g_{\rm B} \, S'_{\rm B}\,,
\end{eqnarray}
where in the last line we define implicitly the scaled
function $S'_{\rm B}$ which is also considered, e.g., in Table 2
of~\cite{JeBeWeSo1999}.
It is a direct consequence of the Dirichlet series (Section 23.2.~of
\cite{AbSt1972})
\begin{equation}
\label{DirichletSer}
\zeta (s) \; = \; \sum_{m=0}^{\infty} \, (m+1)^{-s}\,,
\end{equation}
that we have the inequality $1 \le \zeta(2n+4) \le \zeta(4)$ for all
nonnegative integers $n$. Consequently, the zeta function
does not contribute to the factorial 
divergence of the perturbation series (\ref{Bperser}). 
Thus, the factorial $(2n+1)!$ on the
right-hand side of (\ref{S_B_1}) implies that the perturbation series
for $S_{\mathrm{B}}$ diverges for every 
coupling $g_{\mathrm{B}} \ne 0$. Furthermore, 
it is clear from the representation (\ref{S_B_1}) that an
asymptotically optimized 
$(\kappa, \lambda)$-generalized Borel resummation scheme
for $S_{\mathrm{B}}$ according to (\ref{BorIntTran_1})--(\ref{BorTranSer})
requires the parameter setting $\kappa = \lambda = 2$.

We now discuss the construction
of the asymptotically optimized Borel 
transform explicitly. We start from the scaled series,
\begin{equation}
\label{defSBprime2}
S'_{\rm B}(g_{\mathrm{B}}) \; = \;  
\sum_{n=0}^{\infty} \, (16\,c_n) \, g_{\rm B}^n \; = \;
\sum_{n=0}^{\infty} \, (-1)^n \, (2n+1)! \, 
\frac{2 \zeta(2n+4)}{\pi^{2n+4}} \, g_{\rm B}^n \, .
\end{equation}
The $(2, 2)$-generalized Borel transformed series of
$S'_{\rm B}$ is given by
\begin{equation}
\mathcal{B}^{(2, 2)} \left(S'_{\rm B}; z \right) \; = \;
\sum_{n=0}^{\infty} \, \frac{16\,c_n}{(2n+1)!} \, g_{\rm B}^n \; = \;
\sum_{n=0}^{\infty} (-1)^n \, 
\frac{2 \zeta(2n+4)}{\pi^{2n+4}} \, z^n \, .
\label{DefSBBorTranSer}
\end{equation}
This series can be brought into a form which clearly shows the
singularity structure of the Borel transformed series. For that purpose,
we replace the Riemann zeta function by its Dirichlet series according
to (\ref{DirichletSer}) and interchange the order of the two infinite
nested summations:
\begin{equation}
\label{SupGeom}
\mathcal{B}^{(2, 2)} \left( S'_{\rm B}; z \right) \; = \;
\frac{2}{\pi^4} \, 
\sum_{m=0}^{\infty} \, (m+1)^{-4} \, \sum_{n=0}^{\infty} \, 
\left\{ - z/ \left[ \pi (m+1) \right]^2 \right\}^n \, .
\end{equation}
Thus, $\mathcal{B}^{(2, 2)} \left( S'_{\rm B}; z \right)$ is
essentially a superposition of geometric series with arguments $z/
\left[ \pi (m+1) \right]^2$. 
If we now use $\sum_{k=0}^{\infty} (-x)^k = 1/(1+x)$, we obtain:
\begin{equation}
\label{SBBorTranSer}
\mathcal{B}^{(2, 2)} \left( S'_{\rm B}; z \right) \; = \; 
\frac{2}{\pi^4} \,
\sum_{m=0}^{\infty} \, \frac{(m+1)^{-4}}
{1 + z/ \left[ \pi (m+1) \right]^2} \, .
\end{equation}
This representation shows that the poles of the Borel
transformed series $\mathcal{B}^{(2, 2)} \left( S'_{\rm B}; z
\right)$ are located along the negative real axis according to
\begin{equation}
z \; = \; - n^2 \, \pi^2 \, ,
\end{equation}
where $n$ is a nonzero positive integer. Moreover, we obtain the
following representations for the QED effective action $S_{\mathrm{B}}$
as a $(2, 2)$-generalized Borel integral according to
Eqs.~(\ref{BorIntTran_1}) and (\ref{BorIntTran_2}):
\begin{eqnarray}
\label{AsyOptEffAct}
S_{\mathrm{B}} & = & - \frac{e^2 B^2}{8 \pi^2} \, g_{\rm B} \,
\int_{0}^{\infty} \, t \ 
\mathcal{B}^{(2, 2)} \left( S'_{\rm B}; 
g_{\rm B} t^2 \right) \, \exp (-t) \, \mathrm{d} t
\\
& = & - \frac{e^2 B^2}{8 \pi^2} \, \int_{0}^{\infty} \, s \ 
\mathcal{B}^{(2, 2)} \left( S'_{\rm B}; s^2 \right) \, 
\exp (-s/g_{\mathrm{B}}^{1/2}) \, \mathrm{d} s \, .
\end{eqnarray}

Cveti\v{c} and Yu in~\cite{CvYu2000} use a $(2,2)$-generalized Borel
transformed series for the QED effective action, constructed according
to Eq.~(\ref{AsyOptEffAct}). As explained in Section \ref{Sec:intro},
this transformation is asymptotically optimized in the sense that the
leading factorial growth of the perturbative coefficients in
Eq.~(\ref{S_B_1}) is divided out.  It could appear from the Eqs.~(4) and
(8) in~\cite{CvYu2000} that a $(1,1)$-generalized Borel transform is
used where the $n$th perturbative coefficient is divided by a factor of
$n! = \Gamma(n+1)$. This is, however, not the case. Note that
in~\cite{CvYu2000}, the perturbative expansions are written in a very
peculiar parameter, which is
\begin{equation}
\label{defb}
{\tilde b} = e \, B/m_{\rm e}^2\,.
\end{equation}
In normal QED terminology, this would correspond to an expansion in
powers of $\sqrt{\alpha}$. By consequence, all even-order perturbative
coefficients vanish in the analysis presented in
Ref.~\cite{CvYu2000}. In this context, one may to note that the
expression for the delta transformation according to Eq.~(11)
in~\cite{CvYu2000} is actually undefined since it involves divisions by
zero. The expansion parameter used here is $g_{\rm B} = e^2 \, B^2 /
m_{\rm e}^4 = {\tilde b}^2$; this parameter is also used
in~\cite{JeBeWeSo1999} and in~\cite{DuHa1999}.

Due to the special, mathematically compact form of the perturbative
coefficients in Eq.~(\ref{cn}), the asymptotically optimized Borel
transform (\ref{SupGeom}) is simply a superposition of geometric
series. Such a simple mathematical structure cannot be expected to be of
general importance concerning series occurring in quantum field
theory. Note that this series can be brought in a form which clearly
shows that Wynn's epsilon algorithm, which computes Pad\'e approximants
according to (\ref{Eps_Pade}), is optimal, as discussed in Section
\ref{Sec:ResumMeth}. For that purpose, we rewrite the $n$th partial
sum of the series (\ref{SBBorTranSer}) as follows:
\begin{equation}
\sum_{\nu=0}^{n} \, (-1)^\nu \, \frac{2\,\zeta(2\nu+4)}{\pi^{2\nu+4}} \, 
  z^{\nu} \; = \; 
\mathcal{B}^{(2,2)} (S'_{\rm B}; x) \, - \,
\sum_{\nu=n+1}^{\infty} \, (-1)^\nu \, 
  \frac{2\,\zeta(2\nu+4)}{\pi^{2\nu+4}} \, z^{\nu} \, .
\end{equation}
On substituting the Dirichlet series (\ref{DirichletSer}) into the
infinite series on the right-hand side and interchanging the order of
summations, we obtain
\begin{equation}
\sum_{\nu=0}^{n} (-1)^\nu \, \frac{2\,\zeta(2\nu+4)}{\pi^{2\nu+4}} \, 
  z^{\nu} \; = \; 
\mathcal{B}^{(2,2)} (S'_{\rm B}; z) \, + \,
\frac{2\,(-1)^n}{\pi^4} \, 
\sum_{m=0}^{\infty} \, \frac{z (m+1)^{-6}}{\pi^2 + z/(m+1)^2} \,
\left( \frac{z}{[\pi(m+1)]^2} \right)^n \, .
\end{equation}
Thus, the partial sum of the Borel transformed series
(\ref{SBBorTranSer}) possesses the following general structure:
\begin{equation}
s_n \; = \; s \, + \, (-1)^n \, \sum_{j=0}^{\infty} \, c_j \, 
\lambda_{j}^{n} \, .
\end{equation}
This sequence is obviously a special case of the sequence
(\ref{GenModSeqEps}), for which Wynn's epsilon algorithm is -- as
discussed in Section \ref{Sec:ResumMeth} -- asymptotically optimal.

The conclusions drawn by Cveti\v{c} and Yu in~\cite{CvYu2000} appear to
be restricted, at least in part, to the particular model example studied
in their paper.  In this context it should be emphasized that the
superiority of the delta transformation over Pad\'{e} approximants if
applied {\em directly} to factorially divergent series, cannot be
assumed to persist after the Borel transformation by which the leading
factorial divergence is divided out. I.e., the delta transformation is
more powerful than the Pad\'{e} technique for factorially divergent
series, but this finding by no means allows us to conclude that, or in
fact has any connection to the assumption that the combined Borel-delta
technique should be numerically superior to the Borel-Pad\'{e}
method. This consideration is relevant for the interpretation of the
conclusions drawn by Cveti\v{c} and Yu with regard to the variant of the
Borel method proposed in~\cite{CvYu2000}, which uses the delta
transformation for the analytic continuation of the Borel transformed
series and which is called the Borel-Weniger method by the authors
of~\cite{CvYu2000}.

We return now to the discussion of 
further improvements of the asymptotically
optimized Borel-Pad\'{e} method.
The leading asymptotics do not only permit to modify (optimize) the Borel
transform accordingly, but indeed it is the leading large-order
asymptotics which determine the location of the poles in the Borel
plane. In view of Eq.~(\ref{SBBorTranSer}), the
singularities of the function 
$\mathcal{B}^{(2, 2)} \left( S'_{\rm B}; z \right)$ defined
in (\ref{DefSBBorTranSer}) are at $z = -n^2\, \pi^2$,
i.e.~along the negative real axis.
This is where one would expect them to lie in
(distant) analogy to the renormalon theory~\cite{Be1999}. 

%
%
\begin{table}[htb]     
\begin{center}
\begin{minipage}{12.0cm}
\begin{center}
\caption{\label{table1} Resummation of the divergent series
$S'_{\rm B}$ given in Eq.~(\ref{defSBprime2}) with $g_{\rm B} = 10$.
The numerical data is presented normalized to a number in the
interval $(0,1)$ via multiplication by a factor of $100$.}
\vspace{0.2cm}
\begin{tabular}{ccc}
\hline
\hline
\rule[-0mm]{0mm}{4mm}
$n$ & ${\cal T}'S'_{\mathrm{B},n}$ & ${\cal T}'''S'_{\mathrm{B},n}$ \\
\rule[-0mm]{0mm}{4mm}
 & asymptotically optimized, & improved transforms \\
\rule[-0mm]{0mm}{4mm} 
 & see~Eq.~(\ref{TrafoN}) & 
def.~in Eq.~(\ref{TrafoN3prime}) \\[2ex]
\hline         
\rule[-0mm]{0mm}{5mm} 
0 &
-2.222~222~222~222~222~222 &
-2.222~222~222~222~222~222 \\
\rule[-0mm]{0mm}{5mm} 
1 &
-0.779~860~343~938~511~943 &
-0.846~447~993~882~134~544 \\
\rule[-0mm]{0mm}{5mm} 
2 &
-0.832~545~950~383~972~556 &
-0.807~698~096~764~310~129 \\
\rule[-0mm]{0mm}{5mm} 
3 &
-0.804~166~791~460~607~115 &
-0.805~669~649~913~560~215 \\
\rule[-0mm]{0mm}{5mm} 
4 &
-0.806~776~251~699~410~322 &
-0.805~634~754~493~393~579 \\
\rule[-0mm]{0mm}{5mm} 
5 &
-0.805~531~742~010~943~471 &
-0.805~634~061~009~148~890 \\
\rule[-0mm]{0mm}{5mm} 
6 &
-0.805~700~473~754~628~870 &
-0.805~633~961~318~348~380 \\
\rule[-0mm]{0mm}{5mm} 
7 &
-0.805~626~062~286~745~674 &
-0.805~633~975~558~025~628 \\
\rule[-0mm]{0mm}{5mm} 
8 &
-0.805~638~540~560~183~781 &
-0.805~633~975~330~131~982 \\
\rule[-0mm]{0mm}{5mm} 
9 &
-0.805~633~322~312~161~338 &
-0.805~633~975~322~121~382 \\
\rule[-0mm]{0mm}{5mm} 
10 &
-0.805~634~321~402~303~670 &
-0.805~633~975~321~669~649 \\
\rule[-0mm]{0mm}{5mm} 
11 &
-0.805~633~919~056~749~287 &
-0.805~633~975~321~697~521 \\
\rule[-0mm]{0mm}{5mm} 
12 &
-0.805~634~003~326~479~025 &
-0.805~633~975~321~696~356 \\
\rule[-0mm]{0mm}{5mm} 
13 &
-0.805~633~970~320~418~441 &
-0.805~633~975~321~696~294 \\
\rule[-3mm]{0mm}{8mm} 
14 &
-0.805~633~977~694~044~469 &
-0.805~633~975~321~696~292 \\              
\hline
\rule[-3mm]{0mm}{8mm} 
exact &
-0.805~633~975~321~696~292 &
-0.805~633~975~321~696~292 \\
\hline
\hline
\end{tabular}
\end{center}
\end{minipage}
\end{center}
\end{table}

Using the information on the location of the
poles, it is possible to construct further improved
Pad\'{e} approximants. To this end, we
utilize the known location of the poles in order to
construct improved Pad\'{e}
approximants to the function 
$\mathcal{B}^{(2, 2)} \left( S'_{\rm B}; z \right)$.
Normally, the Borel integral (\ref{SBBorTranSer}) would be evaluated
with the upper- or lower-diagonal Pad\'{e} approximants
to $\mathcal{B}^{(2, 2)} \left( S'_{\rm B}; z \right)$
in the integrand. 
We use upper-diagonal Pad\'{e} approximants here, as they can be
computed by Wynn's epsilon algorithm according to (\ref{Eps_Pade}).
We denote by ${\cal P}_n(z)$ the upper-diagonal Pad\'{e}   
approximant,
\begin{equation}
{\cal P}_n(z) = 
\big[ \trunc{(n+1)/2}/\trunc{n/2}\big]_{\mathcal{B}^{(2, 2)} 
  (S'_{\rm B})}(z)\,.
\end{equation} 
In the upper-diagonal case, 
we evaluate the transforms
${\cal T}S'_{\mathrm{B},n}$ where 
\begin{eqnarray}
\label{TrafoN}
{\cal T}S'_{\mathrm{B},n} & = & 
\int_{0}^{\infty} \, t \ 
{\cal P}_n\left(g_{\rm B} t^2 \right) \, \exp (-t) \, \mathrm{d} t\,,
\end{eqnarray}
and observe numerical convergence of the 
transform at large transformation order $n$.
When the location of the poles is known, we may improve
the convergence of the transforms by the following replacement,
\begin{equation}
\label{poles1}
{\cal P}_n(z) \to {\cal P}'_n(z) =
\frac{{\cal Q}_n(z)}{\prod\limits_{i=1}^{\trunc{n/2}} (1 + z/(n^2\pi^2))}\,,
\end{equation}
where ${\cal Q}_n(z)$ is the $[ \trunc{(n+1)/2} / 0]$-Pad\'{e}
approximant to the function ${\cal R}_n(z)$,
\begin{equation}
\label{poles2}
{\cal Q}_n(z) = \left[ \trunc{(n+1)/2} / 0 \right]_{{\cal R}_n}(z)\,,
\end{equation}
and ${\cal R}_n(z)$ is given by
\begin{equation}
\label{poles3}
{\cal R}_n(z) = \prod\limits_{i=1}^{\trunc{n/2}} (1 + z/(n^2\pi^2)) \,   
  \mathcal{B}^{(2, 2)} \left( S'_{\rm B}; z \right)\,.
\end{equation}
The asymptotic enhancement is possible only if additional asymptotic
information is available on the perturbative coefficients. Such
information may be available (renormalon poles), but this is not
necessarily provided. In this context it should be noted that there is
currently no general proof of the assumption that the renormalon poles
are the only relevant poles in the Borel plane~\cite{Su1999}, but the
factorial divergence of the perturbative coefficients is a commonly
accepted assumption~\cite{Li1977,Li1976Lett,La1977,
ItPaZu1977,BaItZuPa1978,
LeGuZiJu1990,ZJ1996,Be1999}.

The pole-structure improved transforms are obtained from 
(\ref{TrafoN}) by the replacement (\ref{poles1}),
\begin{eqnarray}
\label{TrafoNprime}
{\cal T}'S'_{\mathrm{B},n} & = & 
\int_{0}^{\infty} \, t \ 
{\cal P}'_n\left(g_{\rm B} t^2 \right) \, \exp (-t) \, \mathrm{d} t\,.
\end{eqnarray}
Similar improvement of the convergence of transforms can also be
expected in those cases where the final evaluation of the 
Borel integral proceeds in the complex plane along the
integration contours introduced in~\cite{Je2000}.

Further improvement of the rate of convergence is possible by 
taking the transforms ${\cal T}'S'_{\mathrm{B},n}$ as input data
to the epsilon algorithm (\ref{eps_al}) in order to accelerate
the convergence of the sequence of the pole-structure
improved transforms ${\bf \{} 
{\cal T}'S'_{\mathrm{B},n} {\bf \}}_{n=0}^{\infty}$.
The application of the epsilon algorithm defined
in Eq.~(\ref{eps_al})
to the pole-structure
improved transforms results in a sequence of 
upper-diagonal Pad\'{e} approximants which we denote by
\begin{equation}
\label{TrafoN2prime}
{\cal T}''S'_{\mathrm{B},n}  =
\epsilon_{2 \trunc{n/2}}^{(n - 2 \trunc{n/2})}\,.
\end{equation}
As input data for the epsilon algorithm, we use 
\begin{equation}
s_n = {\cal T}'S'_{\mathrm{B},n}\, .
\end{equation}
In a second epsilon transformation
we may in turn employ the ${\cal T}''S'_{\mathrm{B},n}$ 
as input data for a further application of the epsilon
algorithm,
\begin{equation}
\label{TrafoN3prime}
{\cal T}'''S'_{\mathrm{B},n} = 
\epsilon_{2 \trunc{n/2}}^{(n - 2 \trunc{n/2})}\, ,
\end{equation}
where we use $s_n = {\cal T}''S'_{\mathrm{B},n}$ as input data.
This results in a sequence of pole-structure and doubly
epsilon-improved transforms ${\cal T}'''S'_{\mathrm{B},n}$.
The application of the epsilon algorithm further enhances
the rate of convergence of the pole-structure improved transforms.

In Table~\ref{table1} we present numerical data
for the Borel transforms calculated according
to Eq.~(\ref{TrafoN})  
(in passing we note that
these correspond the method proposed in~\cite{CvYu2000}) 
and the transforms (\ref{TrafoN3prime}).
The first 15 transforms calculated according to Eq.~(\ref{TrafoN})  
exhibit convergence to 8 significant digits,
whereas the pole-structure and epsilon improved
transforms coincide with the exact result to 
within 18 significant digits.

The delta transformation is a general-purpose transformation which has
been proven to be applicable to a {\em wide variety} of alternating
factorially divergent
series~\cite{We1989,Delta1,Delta2,Delta3,WeCiVi1993,We1997,JeBeWeSo1999}.
In the context
of the delta transformation, additional asymptotic information
could be used in order to modify the remainder estimates $\omega_n$
defined in Eq.~(\ref{omegaN})
(see also Eqs.~(7.3-8),~(8.2-7), (8.4-1) and (8.4-4) of~\cite{We1989}). 
Also, we note a rescaling of the perturbative
coefficients as a potential source for further
improvements~\cite{Delta2}.      
Work along these lines is currently in progress and will be 
presented elsewhere~\cite{JeBeMHMoWeSo1999}.

\typeout{::Section 4} 
%
%
\section{ASYMPTOTICALLY OPTIMIZED PREDICTIONS}
\label{Sec:AsyOptPre}

We refer here to the predictions of unknown 
higher-order perturbative coefficients
as perturbative predictions or perturbative extrapolations.
As outlined in Section~\ref{Sec:ResumMeth} and~\cite{SaElKa1995},
these extrapolations are obtained
by reexpanding certain rational approximants in powers of the 
coupling parameter. The next higher-order
term obtained after the reexpansion can then be interpreted as a
prediction for that perturbative coefficient.
The rational
approximants discussed in Section~\ref{Sec:ResumMeth} fulfill
accuracy-through-order relations, i.e., upon reexpansion in the coupling
parameter, all the perturbative terms used for the
construction of the rational approximant are reproduced
[see Eq.~(\ref{O_est_PA}) for Pad\'{e} approximants and
Eq.~(\ref{O_est_delta}) for the delta transformation].
The coefficients of the Borel transformed series
are related to those of the input series by
Eq.~(\ref{BorTranSer}). Therefore, we can either predict the perturbative 
coefficients of the original series or the coefficients of the Borel
transformed series.

We consider here the asymptotic improvement
of three different prediction methods: (i) asymptotically optimized
predictions based on the combination of Borel and
Pad\'{e} techniques, (ii) predictions based on the 
delta transformation, and (iii) the direct application of 
Pad\'{e} approximants to the perturbation series.
We consider here the following improvements beyond
reexpansion of the rational approximant, 
\begin{enumerate}
\item A-posteriori corrections. These are further corrections to the 
perturbative predictions obtained by estimating not only the coefficient,
but also the probable error in making that estimate.
Similar methods in the context of Pad\'{e} approximants
have been introduced, e.g., in~\cite{ChElSt1999a}.
\item Fixing poles. In the
context of the asymptotically improved Borel-Pad\'{e} method, it is possible
to improve predictions if the leading and subleading large-order asymptotics
of the  perturbative predictions are known. These asymptotics determine
the location of the poles in the Borel plane, which can be put in
by hand (see also Eqs.~(\ref{poles1})--(\ref{poles3}) 
in Section~\ref{Sec:AsyOptRes} and, in part,~\cite{Cv2000}).
\item Renormalization group. The renormalization group can be used to
enhance perturbative predictions for certain classes of diagrams;
this has been used e.g.~in~\cite{KaSt1995} for the anomalous magnetic 
moment of the muon.
\end{enumerate}
It is natural to assume that combinations of these techniques
should be investigated where appropriate.
 
The basic idea of a-posteriori corrections is as follows.
The errors made in the ``prediction'' of lower-order
coefficients are available by the time
we come to higher order, so they may be utilized 
for an estimate of the error which is to be expected
in a prediction of the next higher-order coefficient. 
We denote the predictions which are obtained by 
extrapolating the coefficients and the 
``prediction errors'' (in contrast to the coefficients alone)
by the term {\em a-posteriori improved predictions}
because the further correction due to the 
extrapolated error is applied after the reexpansion
of the rational approximant which yields the ``first-order''
prediction. 
The a-posteriori improvement of predictions is useful    
in both lower and higher orders of perturbation theory.
In higher orders, the transient, pre-asymptotic behavior of the 
perturbative coefficients has died away,
and the extrapolations of the coefficients as well as
the a-posterori corrections become more accurate.
A particular merit of the a-posteriori corrections is 
the fact that they can be applied to 
{\em any} of the prediction algorithms proposed above,
in order to achieve an improvement of the prediction beyond the 
reexpansion of the rational approximant used.

%
%
\begin{table}[htb]
\begin{center}
\begin{minipage}{15.0cm}
\caption{\label{table2} Prediction of
perturbative coefficients $\hat{c}_{13}$ 
defined in Eq.~(\ref{defhatcn}).}
\begin{center}
\begin{tabular}{ccc}
\hline
\hline
\rule[-3mm]{0mm}{8mm}
method & result & relative error \\
\hline         
\rule[-0mm]{0mm}{5mm}
Asympt. opt. $[5/6]$-Borel-Pad\'{e} & 
 $2.221~459~447~36 \times 10^{-8}$ &        
 $6 \times 10^{-7}$ \\  
(see~\cite{CvYu2000}) & & \\[2ex]
\rule[-0mm]{0mm}{5mm}
Asympt. opt. $[6/5]$-Borel-Pad\'{e} &
 $2.221~454~724~11 \times 10^{-8}$ &
 $3 \times 10^{-6}$ \\
(for comparison) & & \\[2ex]
\rule[-0mm]{0mm}{5mm}
Asympt. opt. $[6/5]$-Borel-Pad\'{e} &
 $2.221~460~221~71 \times 10^{-8}$ &
 $3 \times 10^{-7}$ \\
with a-posteriori correction & & \\
(this work) & & \\[2ex]
\rule[-0mm]{0mm}{5mm}
Asympt. opt. $[6/5]$-Borel-Pad\'{e} &
 $2.221~460~950~35 \times 10^{-8}$ &
 $3 \times 10^{-8}$ \\
with one pole (this work) & & \\[2ex]
\rule[-0mm]{0mm}{5mm}
Asympt. opt. $[6/5]$-Borel-Pad\'{e} &
 $2.221~460~901~25 \times 10^{-8}$ &
 $1 \times 10^{-8}$ \\
with three poles (this work) & & \\[2ex]
\rule[-0mm]{0mm}{5mm}
Asympt. opt. $[6/5]$-Borel-Pad\'{e} &
 $2.221~460~880~27 \times 10^{-8}$ &
 $5 \times 10^{-10}$ \\
with five poles (this work) & & \\[2ex]
\rule[-0mm]{0mm}{5mm}
Asympt. opt. $[6/5]$-Borel-Pad\'{e} &
 $2.221~460~878~35 \times 10^{-8}$ &
 $3 \times 10^{-10}$ \\
with five poles $+$ a-posteriori (this work) & & \\[2ex]
\hline
\rule[-3mm]{0mm}{8mm}
$\hat{c}_{13}$: exact value &
 $2.221~460~878~99 \times 10^{-8}$ & \\       
\hline
\hline
\end{tabular}
\end{center}
\end{minipage}
\end{center}
\end{table}

We will be investigating in the sequel the Borel transform
of the QED effective action defined in (\ref{SBBorTranSer}). 
In order to investigate
the extrapolation of coefficients of the 
Borel transform, we define 
auxiliary quantities (coefficients) $\hat{c}_{2n+1}$ by
the relation
\begin{equation}
\label{defhatcn}
\hat{c}_{2n+1} = -\frac{16\,c_n}{(2n+1)!}\,,
\end{equation}
where the $c_n$ are defined in Eq.~(\ref{cn}).
We additionally set $\hat{c}_{2n} = 0$
for even-order coefficients.
In terms of the coefficients
$c_j(p)$ which are defined in Eq.~(5) in~\cite{CvYu2000},
the $\hat{c}_{2n+1}$ are given by
$\hat{c}_{2n+1} = (-1)^n \, c_{2n+1}(0)$.
The coefficient $\hat{c}_{2n+1}$, written in
terms of the Bernoulli numbers, reads [see also Eq.~(\ref{cn})]
\begin{equation}
\hat{c}_{2n+1} = (-1)^n \, \frac{2^{2n+4}\,|{\cal B}_{2n+4}|}{(2n+4)!}\,.
\end{equation}

We concentrate here on the coefficient $\hat{c}_{13}$
defined in Eq.~(\ref{defhatcn}) and we discuss how the prediction for the
coefficient $\hat{c}_{13}$ 
can be improved on the basis of a-posteriori
corrections and other asymptotic improvements.
We define {\em correction factors} $\xi_n$ by
\begin{equation}
\hat{c}_n \;\; = \;\; \xi_n \, \bar{c}_n
\end{equation}
where $\hat{c}_n$ is the exact $n$th order coefficient and
$\bar{c}_n$ is the estimate obtained by reexpanding
the rational approximant which is used for the prediction.
In the case of Borel-Pad\'{e} approximants, these would be
Pad\'{e} approximants applied to the Borel transform
of the QED effective action (\ref{SBBorTranSer}).
Specifically, for the prediction of the $n$th perturbative
coefficient, these are the approximants
$\trunc{(n-1)/2} / \trunc{n/2}$ for the lower-diagonal and 
$\trunc{n/2} / \trunc{(n-1)/2}$ for the upper-diagonal case.
For the prediction of $\hat{c}_{13}$, the previous errors
made in the ``prediction'' of  $\hat{c}_{7}$, $\hat{c}_{9}$
and $\hat{c}_{11}$ can be analyzed.
Note that the exact values of  $\hat{c}_{7}$, $\hat{c}_{9}$
and $\hat{c}_{11}$ must be assumed as available, exploitable
information by the time we try to predict $\hat{c}_{13}$.

An estimate for $\xi_{13}$ can be obtained for example
by fitting the natural logarithms
of the quantities $\xi_{7}-1$, $\xi_{9}-1$ and $\xi_{11}-1$ 
with a linear model in order to
obtain an estimate for $\xi_{13}-1$.
This {\em linear} fit of the logarithms
of the $\xi_i$ is based on the 
empirical observation that relative errors of the predictions
decrease exponentially in higher order,
a phenomenon which has been observed in a number of applications,
including variants of anharmonic oscillators.
In the context of Pad\'{e} approximants, a similar error
dependence has been conjectured (see~\cite{ChElSt1999a}). 
The leading coefficient of the 
decay of the relative errors may depend on the
problem considered and on the extrapolation scheme
used, but the exponential improvement of perturbative
predictions in higher order appears to be a rather 
general feature.
Details on this point will be presented
elsewhere~\cite{JeBeMHMoWeSo1999}. 

Using a linear least-squares fit of the respective logarithms
$\ln(\xi_{7}-1)$, $\ln(\xi_{9}-1)$ and $\ln(\xi_{11}-1)$, 
we obtain an estimate of
$\xi_{13} - 1 = 2.47\times10^{-6}$ in the case
of upper-diagonal Borel-Pad\'{e} approximants.
This leads to the data presented in Table~\ref{table2}.
Note that even the crude linear model for the
$\ln(\xi_i - 1)$ used here
already doubles the accuracy of the prediction
of the coefficient $\hat{c}_{13}$ as compared to the 
plain Borel-Pad\'{e} prediction used, for example, in~\cite{CvYu2000}. 
Note also that
an averaging of upper-and lower-diagonal
Borel-Pad\'{e} approximants does not improve
the situation in favor of the plain Borel-Pad\'{e} extrapolation. 
Further improvements of the {\em a-posteriori} corrected 
predictions is possible
with more elaborate extrapolation schemes~\cite{JeBeMHMoWeSo1999}.

We would like to mention that knowledge of the leading 
asymptotics of the perturbative coefficients is required for the construction
of an asymptotically
optimized Borel-Pad\'{e} transformation; this information
is not available for many of the phenomenologically
interesting series currently 
investigated~\cite{ElStChMiSp1998,StEl1998,
ChElMiSt1999,ChElSt1999a,ChElSt1999b,
ChElSt2000,ElChSt2000}.
It is helpful to note that the delta transformation is (like Pad\'{e}) a
rather general-purpose method for the prediction of perturbative
coefficients, and that knowledge of large-order asymptotics is not 
required for its construction or application in a particular case.
This property is helpful especially in cases of practical interest
where little is rigorously known about the large-order
asymptotics of the perturbative coefficients, and where only a 
limited number of perturbative coefficients are available.
A number of practically
interesting examples of delta-based predictions were discussed
in~\cite{JeBeWeSo1999}, and it was observed that 
the delta transformation yields more accurate predictions than the Pad\'{e}
technique in many cases. 

We now turn to a discussion of topologically 
new effects in higher orders of perturbation theory
and improvements of predictions based on the renormalization group.
Topologically new effects have 
caused problems for perturbative predictions
in the past. We refer to the quartic Casimirs
in the QCD beta function~\cite{ElKaSa1997,RiVeLa1997,ElJaJoKaSa1998} and
to light-by-light scattering graphs in the 
tenth order anomalous magnetic moment of the muon~\cite{KaSt1995}.
Analogous considerations might hold for 
the perturbation series investigated in~\cite{Cv2000}.
The topologically new effects
cannot be taken into account by straight extrapolations,
nor by renormalization-group improved extrapolations,
which lead to resummation of certain classes of diagrams.
For the anomalous magnetic moment of the
muon discussed in~\cite{KaSt1995}, the contribution of the topologically new
light-by-light scattering diagrams originally analyzed in~\cite{Ka1993} could 
not be reproduced by renormalization-group
techniques introduced in~\cite{KaSt1995}. 
By reexpansion of the delta approximant to the 
perturbation series for the muon anomaly, an estimate 
of $a^{(10)}_{\mu} = 711$ has been obtained~\cite{heidelberg1999}. 
If an a-posteriori correction based on a combination
of the delta transformation and Pad\'{e} approximants
is added to this prediction,
then the estimate for the 10th order coefficient
changes to $a^{(10)}_{\mu} \sim 970$~\cite{JeBeMHMoWeSo1999}.
This improved estimate is in excellent
agreement with the analytically obtained approximate result of
$930(170)$ from~\cite{Ka1993},
comprising the topogically new effects which are present in five-loop order.

Now it is of course not permissible to conclude that 
topologically new effects can be taken into account
in the general case by considering a-posteriori corrections,
and/or that a-posteriori corrections necessarily 
give an accurate estimate for the size of these problematic,
topologically new effects.
On the other hand, the reverse proposition, which is that 
perturbative predictions are scientifically unsound because
of their general inability to include topologically new effects,
does not appear to be generally valid, either.

\typeout{::Section 5}
%
%
\section{THE ANOMALOUS DIMENSION}
\label{Sec:PerSerAno}

We consider the divergent perturbation series 
for the $\gamma$ function (anomalous dimension) of the 
Yukawa coupling as studied in~\cite{BrKr1999}. Specifically, we
consider the resummation of the perturbation series for the anomalous dimension
of a fermion field with a Yukawa interaction 
$g\,\bar{\psi}\,\sigma\,\psi$ at $d_c = 4$, which is given in
Eq.~(17) in~\cite{BrKr1999}. This calculation
comprises an evaluation of the contribution of all nested self-energy 
diagrams to the anomalous dimension $\gamma$ function of
the Yukawa theory up to 
the 30-loop level
(an analogous analysis is performed in~\cite{BrKr1999}
for the $(4-\epsilon)$--dimensional $\phi^4$ theory). 
We restrict the discussion here to the Yukawa case.
With the convention
\begin{equation}
a = \frac{g^2}{4 \pi^2}\,,
\end{equation}
the result for the anomalous
dimension $\gamma$ function as considered in~\cite{BrKr1999} reads, 
\begin{equation}
\label{DimGammaYuka}
{\tilde \gamma}_{\rm hopf}(a) \sim \sum_{n=1}^{\infty} \,
  (-1)^n\,\frac{{\tilde G}_n}{2^{2n-1}} \,a^n\,.
\end{equation}
The perturbative coefficients ${\tilde G}_n$ are listed in 
Table~\ref{table3}. The coefficients grow factorially
in absolute magnitude; in~\cite{BrKr1999} little change is 
observed in the quantities
\begin{equation}
\label{asymp1}
{\tilde S}_n = \frac{{\tilde G}_n}{2^{n-1}\,\Gamma(n+1/2)}
\end{equation}
for large $n$. 
The evaluation~\cite{BrKr1999} confirms in a concrete, 30-loop calculation
the assumption originally put forth by
Dyson~\cite{Dy1952} that the convergence radius of the quantum field
theoretic perturbative expansion is zero.
For a large number of quantum field
theoretic observables like
the anomalous magnetic moment of the 
muon (see Section~\ref{Sec:AsyOptPre}) 
only a few perturbative terms are known.
Although rapid growth of the 
perturbative coefficients is observed even in 
relatively low order (see the large number of
examples discussed in~\cite{SaElKa1995}),
one may argue that the factorial growth 
of the coefficients has not been demonstrated 
in concrete calculations, and that it is unclear if severe cancellations 
between different classes of diagrams occur in higher order. 

%
%
\begin{table}[htb]     
\begin{center}
\begin{minipage}{11.0cm}
\begin{center}
\caption{\label{table3} Perturbative coefficients 
${\tilde G}_n$ for the anomalous dimension $\gamma$ function of the 
Yukawa coupling [see Eq.~(\ref{DimGammaYuka})].}
\vspace{0.2cm}
\begin{tabular}{cr}
\hline
\hline
\rule[-3mm]{0mm}{8mm}
$n$ & ${\tilde G}_n$ \\[2ex]
\hline         
1 &  1 \\
2 &  1 \\
3 &  4 \\
4 &  27 \\
5 &  248 \\
6 &  2~830 \\
7 &  38~232 \\
8 &  593~859 \\
9 &  10~401~712 \\
10 &  202~601~898 \\
11 &  4~342~263~000 \\
12 &  101~551~822~350 \\
13 &  2~573~779~506~192 \\
14 &  70~282~204~726~396 \\
15 &  2~057~490~936~366~320 \\
16 &  64~291~032~462~761~955 \\
17 &  2~136~017~303~903~513~184 \\
18 &  75~197~869~250~518~812~754 \\
19 &  2~796~475~872~605~709~079~512 \\
20 &  109~549~714~522~464~120~960~474 \\
21 &  4~509~302~910~783~496~963~256~400 \\
22 &  194~584~224~274~515~194~731~540~740 \\
23 &  8~784~041~120~771~057~847~338~352~720 \\
24 &  414~032~133~398~397~494~698~579~333~710 \\
25 &  20~340~342~746~544~244~143~487~152~873~888 \\
26 &  1~039~819~967~521~866~936~447~997~028~508~900 \\
27 &  55~230~362~672~853~506~023~203~822~058~592~752 \\
28 &  3~043~750~896~574~866~226~650~924~152~479~935~036 \\
29 &  173~814~476~864~493~583~374~050~720~641~310~171~808 \\
30 &  10~272~611~586~206~353~744~425~870~217~572~111~879~288 \\
\hline
\hline
\end{tabular}
\end{center}
\end{minipage}
\end{center}
\end{table}

The 30-loop calculation by
Broadhurst and Kreimer may indicate in one particular example 
at least, that the factorial divergence is likely to persist,
and that possible cancellations due to the 
renormalization or between different sets of diagrams do not
contradict the concept of ultimate factorial divergence
of the perturbative coefficients.
From the observation made in~\cite{BrKr1999}
that the $S_n$ defined
in Eq.~(\ref{asymp1}) change little at large $n$,
one may tentatively infer the leading factorial divergence
of the perturbative coefficients,
\begin{equation}
\label{LeadingGn}
{\tilde G}_n \sim 2^{n-1}\,\Gamma(n+1/2) \quad n\to\infty\,.
\end{equation}
This asymptotic behavior, of course, leads to a vanishing radius of convergence
of the perturbative expansion (\ref{DimGammaYuka}).

As observed by Broadhurst and Kreimer, the perturbation series 
(\ref{DimGammaYuka}) 
can be resummed with the help of an asymptotically improved Borel-Pad\'{e} 
technique. From Eq.~(22) in~\cite{BrKr1999}, it is clear
that a $(1,1/2)$-generalized Borel-Pad\'{e} transformation
is used by Broadhurst and Kreimer [for the definition of 
generalized Borel-Pad\'{e} transformations
see Eqs.~(\ref{BorIntTran_1})--(\ref{BorTranSer}) 
in this article].
This is not completely obvious 
because the transformation as used by Broadhurst and Kreimer
has been modified additionally such as to normalize
the first coefficient of the Borel transform 
(not of the input series) to unity,
and the transformation is additionally 
rewritten such as to reflect the vanishing 
coefficient of zeroth order in $a$ in Eq.~(\ref{DimGammaYuka}).
From the leading asymptotics in Eq.~(\ref{LeadingGn}), 
the singularity
of the $(1,1/2)$-generalized Borel transform 
closest to the origin can be inferred.
This singularity was explicitly ``put in by hand''
by Broadhurst and Kreimer (see Eq.~(22) in~\cite{BrKr1999}). 

It is also possible to resum the 
alternating divergent series (\ref{DimGammaYuka})
by a delta transformation, even at large coupling. At a large 
Yukawa coupling of $g = 30$, 
we obtain a relative accuracy of 6 significant figures 
in the resummed results with a plain, unmodified delta transformation. 

We add here a remark
on the relation of the asymptotically optimized Borel-Pad\'{e} based
predictions to those obtained using the delta transformation.
We consider the 
relative accuracy of perturbative predictions 
for the coefficient
${\tilde G}_{30}$ of the perturbation series
defined in Eq.~(\ref{DimGammaYuka}) using various methods.
The coefficient ${\tilde G}_{30}$ is known (see Table~\ref{table3}),
therefore we merely check the accuracy to which this coefficient
can be reproduced by considering the first 29 
perturbative coefficients of the series (\ref{DimGammaYuka}).
With an asymptotically optimized
$(1,1/2)$-generalized Borel-Pad\'{e} technique, 
the coefficient ${\tilde G}_{30}$ can be reproduced 
with a relative accuracy of 
$5\times10^{-16}$. The delta transformation, without any 
modifications, leads to a prediction with a relative 
error of $3\times10^{-16}$; this result
is more accurate than the prediction provided by the
asymptotically optimized Borel-Pad\'{e} transformation.
In accordance with the results of Section~\ref{Sec:AsyOptPre},
the Borel-Pad\'{e} transformation can be significantly enhanced
by including the pole closest to the origin
(see Eq.~(\ref{LeadingGn}) above and Eq.~(22) in~\cite{BrKr1999}). 
When this pole is included,
a prediction is obtained with a relative
error of $4\times10^{-17}$. 
We do not consider a-posteriori improvements to either 
of these predictions, here.

\typeout{::Section 6}
%
%
\section{CONCLUSION}
\label{Sec:Conclu}

We have considered the resummation of divergent perturbation series
and the prediction of unknown higher-order perturbative coefficients
(perturbative predictions or perturbative extrapolations).
We have mentioned and discussed the following resummation prescriptions,
\begin{itemize}
\item the direct application of Pad\'{e} approximants to 
a divergent series,
\item the direct application of the nonlinear (delta) sequence
transformation, 
\item asymptotically improved variants of the Borel-Pad\'{e} technique.
\end{itemize}
The direct application of Pad\'{e} approximants is less efficient
in the resummation of divergent perturbation series than both
the delta transformation and the combined Borel and Pad\'{e} techniques.
The combined, asymptotically improved Borel and Pad\'{e} techniques,
and the delta transformation are complementary. On the one hand,
it can hardly be overemphasized 
that the asymptotically improved Borel-Pad\'{e} technique is less
general than the delta transformation because it
depends on the availability of information on the leading
large-order asymptotics of the coefficients.
By contrast, there is considerable evidence that
the plain, unmodified delta transformation can sum factorially divergent
alternating
series which diverge as strongly as $(3n)!$~\cite{Delta2,WeCiVi1993}. 
This is beyond the power of directly applied 
Pad\'{e} approximants and also beyond the power of the
$(1,1)$-generalized Borel-Pad\'{e} transformation (``usual'')
Borel-Pad\'{e} transformation defined in Eq.~(\ref{BorTranSer}).

If additional information is available on the 
input series, then the asymptotically optimized Borel-Pad\'{e}
technique is rather attractive. As discussed in Section~\ref{Sec:AsyOptRes}, 
it is possible to enhance the rate of convergence simply by
utilizing 
the location of known poles in the Pad\'{e} approximants
to the Borel transform of the input series. These improvements
are not restricted to the first UV and IR renormalon
poles, but, as shown in Eqs.~(\ref{poles1})--(\ref{poles3}) and
exemplified by the numerical
results in Table~\ref{table1}, can take advantage of an 
in principle unlimited number of poles in the Borel plane.
Other techniques for possible improvements of the Borel-Pad\'{e}
algorithm have been described in~\cite{Cv2000}.
Note that it is also possible to generalize the Borel-Pad\'{e} technique
to those cases where there are poles along the positive 
real axis in which case the Borel integral in 
Eq.~(\ref{BorIntTran_1}) is actually
undefined~\cite{Je2000}. Using the special integration
contours in~\cite{Je2000}, it is even possible to derive 
nonperturbative imaginary parts from real, not complex,
perturbative coefficients. Concerning nonperturbative effects in
quantum field theory we also refer to the recent 
investigation~\cite{ElSpXu2000}.

As discussed in Section~\ref{Sec:AsyOptRes}, it is possible
to accelerate the convergence of the resulting Borel-Pad\'{e}
transforms by subsequent application of Wynn's epsilon algorithm
(Borel-Pad\'{e}-Wynn technique). 
These techniques lead to an improved rate of convergence.
Note that the use of explicit information of the location of 
the poles in the Borel plane and the subsequent improvement of the 
convergence of the transforms by Wynn's epsilon algorithm should also
lead to accelerated convergence in the case of the complex
integrations discussed in~\cite{Je2000}.

Because all the resummation prescriptions discussed above
fulfill accuracy-through-order relations, they can be used to predict
perturbative coefficients (we refer to this 
procedure as perturbative predictions or perturbative extrapolations). The
straightforward predictions are obtained by reexpansion of the rational 
approximant in powers of the coupling parameter. That is to say, we consider
here perturbative predictions based on 
\begin{itemize}
\item the reexpansion of Pad\'{e} approximants directly applied to the 
perturbative (input) series,
\item the reexpansion of nonlinear (delta) sequence
transformations directly applied to the 
perturbative (input) series, 
\item and the reexpansion of Pad\'{e} approximants
applied to the asymptotically 
improved Borel transform of the input series. 
\end{itemize}
As it has been demonstrated in~\cite{JeBeWeSo1999}  
and~\cite{CvYu2000},
the predictions based on the delta transformation and on the
combined Borel and Pad\'{e} techniques yield better results
for the perturbative coefficients of the QED effective action
than the Pad\'{e} approximants alone. In~\cite{JeBeWeSo1999}
we also presented a number of more realistic and practically
interesting examples in which the delta transformation 
leads to better predictions than the Pad\'{e} approximants.

Note that
there is currently no general proof of the assumption that the renormalon poles
are the only relevant poles in the Borel plane~\cite{Su1999}, but the
factorial divergence of the perturbative coefficients is a rather
commonly accepted assumption~\cite{Li1977,Li1976Lett,La1977,
ItPaZu1977,BaItZuPa1978,LeGuZiJu1990,ZJ1996,Be1999}.
We should therefore assume that at least asymptotically,
the perturbation series in quantum field theory
approximate factorially divergent series. This is also confirmed
by the concrete 30-loop calculation presented in~\cite{BrKr1999}.
For many factorially divergent series, 
the delta transformation produces better numerical results
than Pad\'{e} approximants (see,  
e.g.,~Ch.~13 of~\cite{We1989}
and \cite{Delta1,Delta2,Delta3,JeMoSoWe1999,WeCiVi1993,We1997}).
Therefore, the delta transformation can be expected
to provide a competitive alternative to Pad\'{e} approximants.   
We again refer to the large number of recent
publications on Pad\'{e}-based extrapolations in quantum field 
theory~\cite{ElStChMiSp1998,StEl1998,
ChElMiSt1999,ChElSt1999a,ChElSt1999b,
ChElSt2000,ElChSt2000,SaLi1991,SaLiSt1993,Sa1994,SaLiSt1994,
SaLi1994a,SaLi1994b,SaElKa1995,Sa1995a,Sa1995b,
SaLiSt1995,ElGaKaSa1996,ElGaKaSa1996a,ElGaKaSa1996b,
Ga1997,ElKaSa1997,SaAbYu1997,JaJoSa1997,Karliner1998,
ElJaJoKaSa1998,BuAbSaStMa1996}.

We consider here mainly the 
following asymptotic improvements of perturbative extrapolations
\begin{itemize}
\item a-posteriori corrections based on estimates not only for 
the next higher-order coefficients, but also for the error which 
is to be expected in the estimation of that coefficient and
\item the use of known (renormalon) poles in order to fix the
denominatior structure of Pad\'{e} approximants in the context of
the Borel-Pad\'{e} method.
\end{itemize}
The general idea of a-posteriori corrections is the following.
In higher orders of perturbation theory a number of lower-order
coefficients are available which may be used in order to construct 
rational approximants. These coefficients can, apart from being useful
for the construction the approximants itself, also be utilized in order to
obtain an estimate for the expected error in the perturbative prediction.
To this end, the extrapolation procedure is applied to the known
lower-order
terms in the perturbation series. A comparison of the ``predictions''
for the known lower-order with their exact results gives 
lower-order correction factors which may be extrapolated to higher order.
This immediately leads to a correction factor for the 
next higher-order perturbative extrapolation (see Section~\ref{Sec:AsyOptPre}).

Using a-posteriori corrections and the pole
structure, we obtain improved results for
the perturbative predictions of the model problem studied
in~\cite{JeBeWeSo1999,DuHa1999,CvYu2000} (see Table~\ref{table2}). 
The a-posteriori corrections also 
lead to an improved estimate for the 
10th order anomalous magnetic moment of the muon and bring 
the a-posteriori corrected prediction in close agreement with an analytically
obtained estimate~\cite{Ka1993}. 
The renormalization-group analysis can lead to a resummation
of certain classes of Feynman diagrams, but it does not lead
to an understanding of topologically new effects which occur in
higher orders of perturbation theory. This phenomenon has lead 
to problems in perturbative predictions in the past, especially in those
cases where these predictions were improved on the basis of a
renormalization group analysis. We
refer to the analysis by Kataev and Starshenko on the 
10th order anomalous magnetic moment of the muon~\cite{KaSt1995} and 
various investigations on the QCD beta 
function~\cite{ElKaSa1997,ElJaJoKaSa1998}.
Notably, as discussed in Section~\ref{Sec:AsyOptPre},
the a-posteriori corrected prediction 
appears to be consistent with the topologically new effects observed 
in 10th order of perturbation theory and calculated approximately
in~\cite{Ka1993}. 

We would like to stress here again that the concept of a-posteriori
corrections is rather general and can be applied to all prediction algorithms
mentioned above. Specifically, we refer to the investigation
\cite{ChElSt1999a} on significant improvements which can be achieved
in the context of Pad\'{e}-based predictions with this technique. 
The reduction of the magnitude of the a-posteriori correction is an attractive
feature of the predictions based on the delta transformation.

In Sections~\ref{Sec:PerSerAno} we show that for the 
more realistic 30-loop series calculated by Broadhurst and Kreimer 
in~\cite{BrKr1999}, even the asymptotically optimized Borel-Pad\'{e} technique
cannot quite match the accuracy obtainable by the plain delta transformation.
It is only when an additional pole is explicitly put in by hand
in the Pad\'{e} transformations that the combined, asymptotically improved
Borel-Pad\'{e} technique becomes more accurate than the 
plain delta transformation. We stress here that the construction of
the asymptotically improved Borel-Pad\'{e} technique requires in itself
a knowledge of the large-order asymptotics of the perturbative coefficients.
Such information is not available in general cases. Specifically,
in those cases where only a
small number of coefficients are known the leading asymptotics cannot be
reliably inferred from empirical approaches, either.

It has been the purpose of this article to clarify how
resummation algorithms and perturbative extrapolations can be improved
if additional information is available on a particular input series.
We have explained in Sections~\ref{Sec:AsyOptRes}--\ref{Sec:PerSerAno} various
algorithms by which the resummation of divergent series and the prediction of
perturbative coefficients can be improved on the basis of additionally
available asymptotic information on a given series; these improvements can be
applied to the Borel-Pad\'{e} based techniques and to the delta transformation
techniques~\cite{JeBeWeSo1999}.

%
%

\section*{ACKNOWLEDGMENTS}

U.J. acknowldges helpful conversations with P.J. Mohr.
E.J.W. acknowledges support from the
Fonds der Chemischen Industrie. G.S. acknowledges continued support
from BMBF, GSI and DFG.     
  

\begin{thebibliography}{99}

\bibitem{LeGuZiJu1990}
J.~C.\ Le Guillou and J.\ Zinn-Justin (eds.), {\it Large-Order Behaviour
of Perturbation Theory} (North-Holland, Amsterdam, 1990). 

\bibitem{BaGr1996}
G.~A. Baker, Jr., and P. Graves-Morris, {\em Pad\'{e} Approximants}, 
  2nd ed. (Cambridge University Press, Cambridge, 1996).

\bibitem{Bor1899} 
E.\ Borel, Ann.\ sci. Ec.\ norm.\ sup.\ Paris {\bf 16}, 9 (1899).

\bibitem{Bor1928}
E.\ Borel, {\it Le{\c c}ons sur les S\'eries Divergentes}, Second
Edition (Gautier-Villars, Paris, 1928).  Reprinted by \'{E}ditions
Jacques Gabay, Paris, 1988. Translated by C.~L.\ Critchfield and A.\
Vakar, {\it Lectures on divergent series} (Los Alamos Scientific
Laboratory, Translation LA-6140-TR, 1975).

\bibitem{We1989} 
E.~J.\ Weniger, Comput.\ Phys.\ Rep.\ {\bf 10}, 189 (1989).

\bibitem{Delta1}  
D.\ Roy, R.\ Bhattacharya, and S.\ Bhowmick, Comput.\ Phys.\ Commun.\
{\bf 93}, 159 (1996);
R.\ Bhattacharya, D.\ Roy, and S. Bhowmick, Comput.\ Phys.\ Commun.\
{\bf 101}, 213 (1996);
D.\ Roy, R.\ Bhattacharya, and S.\ Bhowmick, Comput.\ Phys.\ Commun.\
{\bf 113}, 131, (1998);
A.\ Sarkar, D.\ Sen, S.\ Haldar, and D.\ Roy, Mod.\ Phys.\ Lett.\ B {\bf
12}, 639 (1998).

\bibitem{Delta2}
E.~J. Weniger, Int.\ J.\ Quantum Chem. {\bf 57}, 265 (1996), Erratum,
Int.\ J.\ Quantum Chem.\ {\bf 58}, 319 (1996);
E.~J. Weniger, Ann. Phys. (N. Y.) {\bf 246}, 133 (1996);
E.~J. Weniger, Phys. Rev. Lett. {\bf 77}, 2859 (1996).

\bibitem{Delta3} E.~J. Weniger, Comput. Phys. {\bf 10},  496 (1996);
U.~D.\ Jentschura, P.J.\ Mohr, and G.\ Soff, Phys.\ Rev.\ Lett.\ {\bf
82}, 53 (1999).

\bibitem{JeMoSoWe1999}
U.~D.\ Jentschura, P.J.\ Mohr, G.\ Soff, and E.~J.\ Weniger, Comput.\
Phys.\ Commun.\ {\bf 116}, 28 (1999).

\bibitem{WeCiVi1993}
E.~J.\ Weniger, J.\ {\v C}{\' \i}{\v z}ek, and F.\ Vinette, J.\ Math.\
Phys.\ {\bf 34}, 571 (1993).

\bibitem{We1997}
E.~J.\ Weniger, Phys.\ Rev.\ A {\bf 56}, 5165 (1997).

\bibitem{JeBeWeSo1999}
U. Jentschura, J. Becher, E. Weniger, and G. Soff, Los Alamos preprint
  hep-ph/9911265.

\bibitem{GrafGrec1978}
S.\ Graffi and V.\ Grecchi, J.\ Math.\ Phys.\ {\bf 19}, 1002 (1978).

\bibitem{Gil1973} 
J.\ Gilewicz, in {\it Pad\'{e} Approximants and Their Applications},
edited by P.~R.\ Graves-Morris (Academic Press, London, 1973), p.\ 99.

\bibitem{ElStChMiSp1998}
V. Elias, T.~G. Steele, F. Chishtie, R. Migneron, and K. Sprague, Phys. Rev. D
  {\bf 58},  116007  (1998).

\bibitem{StEl1998}
T.~G. Steele and V. Elias, Mod. Phys. Lett. A {\bf 13},  3151  (1998).

\bibitem{ChElMiSt1999}
F.~A. Chishtie, V. Elias, V.~A. Miransky, and T.~G. Steele, Los Alamos preprint
  hep-ph/9905291.

\bibitem{ChElSt1999a}
F. Chishstie, V. Elias, and T.~G. Steele, Phys. Lett. B {\bf 446},  267
  (1999).

\bibitem{ChElSt1999b}
F. Chishstie, V. Elias, and T.~G. Steele, Phys. Rev. D {\bf 59},  105013
  (1999).

\bibitem{ChElSt2000}
F. Chishstie, V. Elias, and T.~G. Steele, J. Phys. G {\bf 26},  93  (2000).

\bibitem{ElChSt2000}
V. Elias, F.~A. Chishtie, and T.~G. Steele, Los Alamos Preprint hep-ph/0004140.

\bibitem{SaLi1991}
M.~A. Samuel and G. Li, Phys. Rev. D {\bf 44},  3935  (1991).

\bibitem{SaLiSt1993}
M.~A. Samuel, G. Li, and E. Steinfelds, Phys. Rev. D {\bf 48},  869  (1993).

\bibitem{Sa1994}
M.~A. Samuel, Int. J. Theor. Phys. {\bf 33},  1461  (1995).

\bibitem{SaLiSt1994}
M.~A. Samuel, G. Li, and E. Steinfelds, Phys. Lett. B {\bf 323},  188  (1994).

\bibitem{SaLi1994a}
M.~A. Samuel and G. Li, Phys. Lett. B {\bf 331},  114  (1994).

\bibitem{SaLi1994b}
M.~A. Samuel, G. Li, and E. Steinfelds, Phys. Lett. B {\bf 323},  188  (1994).

\bibitem{SaElKa1995}
M.~A. Samuel, J. Ellis, and M. Karliner, Phys. Rev. Lett. {\bf 74},  4380
  (1995).

\bibitem{Sa1995a}
M.~A. Samuel, Int. J. Theor. Phys. {\bf 34},  903  (1995).

\bibitem{Sa1995b}
M.~A. Samuel, Int. J. Theor. Phys. {\bf 34},  1113  (1995).

\bibitem{SaLiSt1995}
M.~A. Samuel, G. Li, and E. Steinfelds, Phys. Rev. E {\bf 51},  3911  (1995).

\bibitem{ElGaKaSa1996}
J. Ellis, E. Gardi, M. Karliner, and M.~A. Samuel, Phys. Lett. B {\bf 366},
  268  (1996).

\bibitem{ElGaKaSa1996a}
J. Ellis, E. Gardi, M. Karliner, and M.~A. Samuel, Phys. Lett. B {\bf 366},
  268  (1996).

\bibitem{ElGaKaSa1996b}
J. Ellis, E. Gardi, M. Karliner, and M.~A. Samuel, Phys. Rev. D {\bf 54},  6986
   (1996).

\bibitem{Ga1997}
E. Gardi, Phys. Rev. D {\bf 56},  68  (1997).

\bibitem{ElKaSa1997}
J. Ellis, M. Karliner, and M.~A. Samuel, Phys. Lett. B {\bf 400},  176  (1997).

\bibitem{SaAbYu1997}
M.~A. Samuel, T. Abraha, and J. Yu, Phys. Lett. B {\bf 394},  165  (1997).

\bibitem{JaJoSa1997}
I. Jack, D.~R.~T. Jones, and M.~A. Samuel, Phys. Lett. B {\bf 407},  143
  (1997).

\bibitem{Karliner1998}
M. Karliner, Acta Phys. Polon. B {\bf 29},  1505  (1998).

\bibitem{ElJaJoKaSa1998}
J. Ellis, I. Jack, D.~R.~T. Jones, M. Karliner, and M.~A. Samuel, Phys. Rev. D
  {\bf 57},  2665  (1998).

\bibitem{BuAbSaStMa1996}
P. Burrows, T. Abraha, M. Samuel, E. Steinfelds, and H. Masuda, Phys. Lett. B
  {\bf 272},  223  (1996).

\bibitem{CvKo1998}
G. Cveti\v{c} and R. K\"{o}gerler, Nucl. Phys. B {\bf 522},  396  (1998).

\bibitem{Cv1998a}
G. Cveti\v{c}, Phys. Rev. D {\bf 57},  R3209  (1998).

\bibitem{Cv1998b}
G. Cveti\v{c}, Nucl. Phys. B {\bf 517},  506  (1998).

\bibitem{CvYu2000}
G. Cveti\v{c} and J.~Y. Yu, Los Alamos preprint hep-ph/0003241.

\bibitem{We2000} E.~J. Weniger, Los Alamos preprint math.CA/0002111.

\bibitem{GraGreSim1970}
S.\ Graffi, V.\ Grecchi, and B.\ Simon, Phys.\ Lett.\ B {\bf 32}, 631
(1970).
 
\bibitem{Fi1997polo}
J. Fischer, Acta Phys. Polon. B {\bf 27},  2549  (1997).
 
\bibitem{Fi1997}
J. Fischer, Int. J. Mod. Phys. A {\bf 12},  3625  (1997).
 
\bibitem{Fi1999}
J. Fischer, Rep. Math. Phys. {\bf 43},  109  (1999).
 
\bibitem{CaFi1999}
I. Caprini and J. Fischer, Phys. Rev. D {\bf 60},  054014  (1999).
 
\bibitem{CaFi2000}
I. Caprini and J. Fischer, Los Alamos
  preprint hep-ph/0002016.

\bibitem{West1991}
G. West, Phys. Rev. Lett. {\bf 67}, 1388 (1991).

\bibitem{West1999}
G. West, Los Alamos Preprint hep-ph/9911416.
  
\bibitem{DuHa1999}
G.~V. Dunne and T.~M. Hall, Phys. Rev. D {\bf 60},  065002  (1999).  

\bibitem{Sc1951}
J. Schwinger, Phys. Rev. {\bf 82},  664  (1951).  

\bibitem{Je2000}
U.~D. Jentschura, Los Alamos preprint hep-ph/0001135.

\bibitem{Car1926}
T.\ Carleman, {\it Les Fonctions Quasi-Analytiques}, (Gautiers Villars,
Paris, (1926)).

\bibitem{ReSi1978}
M. Reed and B. Simon, {\em Methods of Modern Mathematical Physics IV: 
  Analysis of Operators} (Academic Press, New York, 1978).

%
%

\bibitem{ShaWat1994}
B.\ Shawyer and B.\ Watson, {\it Borel's Method of Summability}
(Oxford. U.P., Oxford, (1994)).

\bibitem{SteSha1996}
B.~Yu.\ Sternin and V.~E.\ Shatalov, {\it Borel-Laplace
Transform and Asymptotic Theory} (CRC Press, Boca Raton, (1996)).

\bibitem{Raczka1991}
P. A. R\c{a}czka, Phys. Rev. D, {\bf 43}, R9 (1991).

\bibitem{Pi1999}
M. Pindor, Los Alamos preprint hep-th/9903151.

\bibitem{CuyWuy1987}
A.\ Cuyt and L.\ Wuytack, {\it Nonlinear Methods in Numerical Analysis}
(North-Holland, Amsterdam, 1987).

\bibitem{Wy1956}
P.\ Wynn, Math.\ Tables Aids Comput.\ {\bf 10}, 91 (1956).

\bibitem{Ba1965}
G.~A.\ Baker, Jr.,  Adv.\ Theor.\ Phys.\ {\bf 1}, 1 (1965). 

\bibitem{Bre1991}
C.\ Brezinski, {\it History of Continued Fractions and Pad\'e
Approximants} (Springer-Verlag, Berlin, 1991).

\bibitem{BreRZa1991} 
C.\ Brezinski and M.\ Redivo Zaglia, {\it Extrapolation Methods\/}
(North-Holland, Amsterdam, 1991).

\bibitem{Wy1966}
P.\ Wynn, SIAM J.\ Numer.\ Anal.\ {\bf 3}, 91 (1966).

\bibitem{Lev1973} 
D.\ Levin, Int.\ J.\ Comput.\ Math. B {\bf 3}, 371 (1973).

\bibitem{SmiFo1979}
D.~A. Smith and W.~F. Ford, Math.\ Comput.\ {\bf 38}, 481 (1982).

\bibitem{WeCi1990} 
E.~J.\ Weniger and J.\ {\v C}{\' \i}{\v z}ek, Comput.\ Phys.\ Commun.\
{\bf 59}, 471 (1990).

\bibitem{We2000a} 
E.J.\ Weniger, Comput.\ Phys.\ Commun., submitted. Los Alamos preprint
math.NA/0003227.

%
%

\bibitem{CvYu1999}
G. Cvetic and J.~Y. Yu, Los Alamos preprint hep-ph/9911376.

\bibitem{ItZu1980}
C. Itzykson and J.~B. Zuber, {\em Quantum Field Theory} (McGraw-Hill, 
  New York, NY, 1980).  

\bibitem{AbSt1972}
M. Abramowitz and I.~A. Stegun, {\em Handbook of Mathematical Functions}, 10
  ed. (National Bureau of Standards, Washington, D. C., 1972).

\bibitem{Be1999}
M. Beneke, Phys. Rep. {\bf 317},  1  (1999).

\bibitem{BrKr1999}
D. Broadhurst and D. Kreimer, Los Alamos preprint hep-ph/9912093.

\bibitem{Su1999}
I.~M. Suslov, Zh. \'{E}ksp. Teor. Fiz. {\bf 116},  369  (1999), [JETP {\bf 89},
  197 (1999)].

\bibitem{Li1976Lett}
L.~N. Lipatov, Zh. \'{E}ksp. Teor. Fiz. Pis'ma {\bf 24},  179  (1976), [JETP
  Lett. {\bf 24}, 157 (1976)].

\bibitem{Li1977}
L.~N. Lipatov, Zh. \'{E}ksp. Teor. Fiz. {\bf 72},  411  (1977), [JETP {\bf 45},
  216 (1977)].

\bibitem{La1977}
B. Lautrup, Phys. Lett. B {\bf 69},  109  (1977).

\bibitem{ItPaZu1977}
C. Itzykson, G. Parisi, and J.~B. Zuber, Phys. Rev. D {\bf 16},  996  (1977).

\bibitem{BaItZuPa1978}
R. Balian, C. Itzykson, J.~B. Zuber, and G. Parisi, Phys. Rev. D {\bf 17},
  1041  (1978).

\bibitem{ZJ1996}
J. Zinn-Justin, {\em Quantum Field Theory and Critical Phenomena}, 3rd ed.
  (Clarendon Press, Oxford, 1996).

\bibitem{JeBeMHMoWeSo1999}
U. Jentschura, J. Becher, M. Meyer-Hermann, P.~J. Mohr, E. Weniger, and G.
  Soff, Resummation of the Derivative Expansion of the QED Effective
  Action, {\em in preparation}.

\bibitem{heidelberg1999}
U. D. Jentschura, J. Becher, G. Soff, P. J. Mohr and E. J. Weniger,
Talk given at the workshop ``Lepton Moments 99'' in Heidelberg, Germany (1999),
transparencies available at the URL: 
http://www.physi.uni-heidelberg.de/\~{}muon/lep/lep.html.

\bibitem{Dy1952}
F.~J. Dyson, Phys. Rev. {\bf 85},  631  (1952).

\bibitem{RiVeLa1997}
T. van Ritbergen, J. A. M. Vermaseren and S. A. Larin,
Phys. Lett. B {\bf 400}, 379 (1997).

\bibitem{KaSt1995}
A. L. Kataev and V. S. Starshenko, Phys. Rev. D {\bf 52},
  402 (1995).             

\bibitem{Ka1993}
S. G. Karshenboim, Yad. Fiz. {\bf 56},
  115 (1993) [Phys. At. Nucl. {\bf 56}, 777 (1993)].

\bibitem{Cv2000}
G. Cveti\v{c}, Los Alamos preprint hep-ph/0003123.

%

\bibitem{ElSpXu2000}
V. Elias, K. B. Sprague and Y. Xue, Found. Phys. {\bf 30}, 439 (2000).


\end{thebibliography}
\end{document}